\documentclass[%
 reprint,
nofootinbib,
 amsmath,amssymb,
 aps,
 prd,
 floatfix,
]{revtex4-2}
\usepackage{graphicx}
\usepackage{dcolumn}
\usepackage{bm}

\usepackage{natbib}
\usepackage{latexsym} 
\usepackage{times} 
\usepackage{hyperref}
\usepackage{mathrsfs}
\usepackage[utf8x]{inputenc}
\usepackage{tikz}
\usepackage{booktabs}
\usepackage{array}
\usepackage{multirow}
\usepackage{mathtools}
\usepackage{cancel}
\usepackage{comment}
\usepackage{float}
\usepackage[normalem]{ulem}

\newcommand{\citeAliasTwo}[2]{\defcitealias{#1}{#2}\citetalias{#1}}

\begin{document}

\title{Fully General Relativistic Magnetohydrodynamic Simulations of
Accretion Flows onto Spinning Massive Black Hole Binary Mergers}
\author{Federico Cattorini$^{1,2}$}
 \email{fcattorini@uninsubria.it}
\author{Bruno Giacomazzo$^{3,2,4}$}
\author{Francesco Haardt$^{1,2,4}$}
\author{Monica Colpi$^{3,2}$}
 \affiliation{$^1$Dipartimento di Scienza e Alta Tecnologia, Universit\'a degli Studi dell'Insubria, Via Valleggio 11, I-22100, Como, Italy}
\affiliation{%
 $^2$INFN, Sezione di Milano-Bicocca, Piazza della Scienza 3, I-20126 Milano, Italy}%
 \affiliation{%
 $^3$Dipartimento di Fisica G. Occhialini, Universit\`a di Milano-Bicocca, Piazza della Scienza 3, I-20126 Milano, Italy}%
 \affiliation{%
 $^4$INAF, Osservatorio Astronomico di Brera, Via E. Bianchi 46, I-23807 Merate, Italy}%

\date{\today}

\begin{abstract}
We perform the first suite of fully general relativistic magnetohydrodynamic simulations of spinning massive black hole binary mergers. We consider binary black holes with spins of different magnitudes aligned to the orbital angular momentum, which are immersed in a hot, magnetized gas cloud. We investigate the effect of the spin and degree of magnetization (defined through the fluid parameter $\beta^{-1}\equiv p_{\mathrm{mag}}/p_{\mathrm{fluid}}$) on the properties of the accretion flow. We find that magnetized accretion flows are characterized by more turbulent dynamics, as the magnetic field lines are twisted and compressed during the late inspiral. Pos-merger, the polar regions around the spin axis of the remnant Kerr black hole are magnetically dominated, and the magnetic field strength is increased by a factor $\sim$10$^2$ (independently from the initial value of $\beta^{-1}$). The magnetized gas in the equatorial plane acquires higher angular momentum, and settles in a thin circular structure around the black hole. We find that mass accretion rates of magnetized configurations are generally smaller than in the unmagnetized cases by up to a factor $\sim$3.
Black hole spins have also a suppressing effect on the accretion rate, as large as $\sim$48\%. As a potential driver for electromagnetic emission we follow the evolution of the Poynting luminosity, which increases after merger up to a factor $\sim2$ with increasing spin, regardless of the initial level of magnetization of the fluid. Our results stress the importance of taking into account both spins and magnetic fields when studying accretion processes onto merging massive black holes.
\end{abstract}

\maketitle

\section{Introduction}
Massive black hole binary (MBHB) mergers are natural outcome of galaxy collisions \cite{Begelman-1980, Kormendy-2013}, and are among the most powerful sources of gravitational waves (GWs) which will be detected by future space-based interferometers such as LISA \cite{LISA-2017}. These mergers may occur in gas-rich environment \cite{Barnes-1992, Barnes-1996,Mayer2007, Dotti-2009, Dotti-2012,Chapon2013, Colpi-2014}, leading to the intriguing possibility of concurrent electromagnetic (EM) emission observable by traditional astronomical facilities. Observing these powerful events both in the EM and the GW windows will provide unique opportunities for multimessenger astronomy. 
A major goal of forthcoming multiband EM observations (e.g., {\it Athena} \cite{McGee-2020}) is to observe and study the EM counterparts to LISA MBHB coalescences: detecting the EM signal emitted alongside an ongoing merger will let us probe the existence of multiple disk structures around the massive black holes (MBHs)  \cite{Haiman-2009,Tang-2018,Bowen-2017,Bowen-2018} and, possibly, the launch of  relativistic jets during the inspiral and when the new MBH has formed \cite{Palenzuela-2010,Khan-2018}.
Concurrent observation of EM counterparts to GW events will help illuminate  the physical processes that power quasars, and offer new opportunities for testing the propagation of GWs on cosmological scales, e.g., measuring the differences in the arrival times of light and GWs, or inferring the redshift $z$ versus luminosity distance $d_L$ relation without resorting to EM distance scale calibrators \cite{Schutz-nature,Koksic-2008,Tamanini-2016}.

Our knowledge of the properties of the EM signals emerging during a MBHB merger is still incomplete, despite recent advances \cite{Roedig-2014,Kelly-2017, Tang-2018,D'Ascoli-2018,Yuan2021}.  Predictions on this EM emission depend on the fueling rate; on the hydrodynamical, geometrical and radiative properties of the accreting magnetized gas; and on the MBHs masses and spins.

The development of numerical relativity (NR) simulations of these powerful events is required to advance our theoretical understanding of the physical mechanisms which drive EM signals associated to GW detections. A jump in the predictive power of NR simulations will allow  better predictions for the EM spectrum rising during the late inspiral and coalescence of MBHBs, providing guidance to future observations and maximizing the scientific return of LISA.

The structure of the accretion flows around coalescing MBHBs largely depends on the angular momentum content of the accreting gas conveyed in the galactic merger, and on its thermodynamical state. Two limiting scenarios bracket the range of physical properties of accreting fluids around  MBHBs:
\begin{enumerate}
 \item[(i)] The \textit{circumbinary disk} (CBD) model, in which a  rotationally supported disk surrounds the binary, and  viscous and gravitational torques balance to clear a central cavity at twice the MBHB separation \cite{Milosavljevic-2005}. Numerical simulations show that the system evolves into a nonaxisymmetric configuration  with the cavity becoming highly lopsided and filled with 
 a tenuous, shocked plasma, in part ejected against the disk wall where it loses angular momentum to feed the MBHs. This leads to the formation of two narrow streams, which periodically convey mass onto the MBHs in the form of transient ``minidisks'' that persist down to coalescence
 \citep[e.g.,][]{Noble-2012, D'Orazio-2013, Farris-2014a, Farris-2015b,Tang-2017,Tang-2018,Bowen-2018,Bowen-2019}.
 The first simulations of equal-mass, nonspinning binaries in \textit{magnetized} CBDs were performed by \cite{Noble-2012} (adopting high-order PN approximations) and \cite{Farris-2012} (in full general relativity).
 
 \item[(ii)] If the surrounding gas is hot, tenuous, and not rotationally supported, the MBHs may find themselves embedded in a turbulent and radiatively inefficient accretion flow \cite{Ichimaru-1977, Narayan-1994}. In this scenario, the gas is unable to cool efficiently, and thus the energy is stored in the accretion flow as thermal energy instead of being radiated.  We refer to this scenario as the \textit{gas cloud} model \cite{Farris-2010, Bode-2010, Bode-2012}. The first general relativistic hydrodynamical simulations of merging equal-mass binaries in unmagnetized gas clouds were carried out by  Refs. \cite{Farris-2010} and \cite{Bode-2010} (the latter considered both nonspinning and parallel-spin binaries). These works established that the phases of late inspiral and merger are accompanied by a gradual rise in the emitted bremsstrahlung luminosity, followed by a sudden dropoff corresponding to the postmerger accretion of the shock-heated gas. In a subsequent work \cite[][]{Bode-2012}, the impact of misaligned spins and unequal mass ratios on the physics of hot accretion flows was investigated, and it was found that less symmetric systems result in lower luminosity and delayed emission from the regions near the BHs.
\end{enumerate}

In the present work we consider the hot gas cloud model. We perform the first general relativistic magnetohydrodynamic (GRMHD) simulations of merging spinning BHs immersed in an initially homogeneous fluid, and examine how magnetic fields and spins affect the dynamics of the gas and the Poynting luminosity emission. Our simulations revise the scenario analyzed in Giacomazzo et al. \cite{Giacomazzo-2012} (\citeAliasTwo{Giacomazzo-2012}{Gi12} hereafter) and Kelly et al.  \cite{Kelly-2017} (\citeAliasTwo{Kelly-2017}{Ke17} hereafter), and explore the behavior of moderately magnetized accretion flows (MMAFs) onto binaries of MBHs within the ideal MHD limit. The analysis of moderately magnetized plasma bridges the study of unmagnetized gaseous environments \cite[e.g., ][]{Bode-2010, Farris-2010} and results obtained in the force-free  regime \cite[e.g., ][]{Palenzuela-2010}, which approximate magnetically dominated plasma (i.e., fluids for which $\beta^{-1} \equiv p_{\mathrm{mag}} / p_{\mathrm{fluid}} \gg 1$).

The simulations of \citeAliasTwo{Giacomazzo-2012}{Gi12} were the first to study the nature of MMAFs around equal-mass, nonspinning black hole binaries (set at an initial separation of 8.48 $M$, where $M$ is the total mass of the binary), solving the GRMHD equations with the \texttt{WhiskyMHD} code \cite[][]{Giacomazzo-2007, Giacomazzo-2011}. \citeAliasTwo{Giacomazzo-2012}{Gi12} considered two models for the gas cloud surrounding the binary, both with an initially uniform rest-mass density $\rho_0$: a unmagnetized plasma, and a plasma threaded by an initially uniform magnetic field with an initial ratio of magnetic-to-fluid pressure $\beta^{-1}$ equal to 0.025. Their results showed that MMAFs exhibit different dynamics compared to unmagnetized accretion flows, and can lead to strong, collimated EM emission. 

The results of \citeAliasTwo{Giacomazzo-2012}{Gi12} were farther extended by \citeAliasTwo{Kelly-2017}{Ke17}, who covered a broader collection of physical scenarios adopting the \texttt{IllinoisGRMHD} code \cite[][]{Noble-2006, Etienne-2015} to solve the GRMHD equations. The simulations of \citeAliasTwo{Kelly-2017}{Ke17} consider equal-mass binaries of nonspinning black holes with initial separations covering values between $6.6M$ and $16.3M$. Evolving higher-separation binaries allowed them to better resolve the timing features of the EM (Poynting) emission. Several configurations differing only in the initial magnetic field value $b_0$ were also evolved, showing that the level of the Poynting luminosity reached during the inspiral is little sensitive to the initial magnetic field strength.

In this work, we progress studying the scenario examined by \citeAliasTwo{Giacomazzo-2012}{Gi12}-\citeAliasTwo{Kelly-2017}{Ke17} and carry out the first 3-dimensional GRMHD simulations of merging BHs including spins.
Extending the study to BHs with nonzero spin is key when considering binaries of massive black holes. The motivation is astrophysical, as there is observational evidence that MBHs have grown primarily by efficient accretion \citep{Marconi2004}, and are expected to acquire a non-vanishing spin depending on whether accretion is prograde or retrograde, coherent or chaotic, as discussed extensively in the literature \cite[see, e.g., Refs. ][]{Gammie-2004, King2005,Berti-Volonteri-spin-2008,Dotti-spin02013,Sesana2014, Izquierdo-Villalba-2021}. 

Two recent works presented preliminary results describing GRMHD premerger simulations of spinning binary black holes. Lopez Armengol et al. \cite{Lopez-Armengol-2021} construct an approximate space-time metric, which they name ``Superimposed Kerr-Schild'' (SKS) metric, and carry out simulations of circumbinary accretion onto binary systems with separation fixed at $20M$, and spin parameters $a=(0, \pm 0.9)$. They find that spin can significantly affect the circumbinary accretion via frame-dragging effects, enhancing or reducing it according to the sign of the spin-orbit coupling. Paschalidis et al. \cite{Paschalidis-2021} perform fully general relativistic MHD simulations of BHBs, and consider binary configurations of spinning BHs set at an initial distance $d=20M$, with spin parameters $a=(0, \pm 0.75)$. Their work addresses the formation and dynamics of minidisks. In particular, they demonstrate the impact of spin in allowing the formation of minidisks.
Both aforementioned investigations focus on the late stages of BHB inspiral, i.e. the premerger phase.
By contrast, our work considers different environment (i.e., the gas cloud model) and examines \textit{merging} binary systems, treating both pre- and postmerger phases.

Our simulations consider binary equal-mass BHs with equal spins, both aligned with the orbital angular momentum, and with spin dimensionless parameters of magnitude $a_{1,2}=(0, 0.3, 0.6)$, immersed in a uniform plasma with different initial degrees of magnetization. 
The binary evolutions are carried out with the \texttt{Einstein Toolkit}\footnote{\url{http://einsteintoolkit.org}} \cite[][]{Loffler-2012} on adaptive-mesh refinement (AMR) grids provided by the Carpet driver \cite[][]{Schnetter-2004}. The space-time metric evolution is obtained using the Kranc-based \texttt{McLachlan} \cite[][]{Husa-2006, Brown-2009} thorn, adopting the BSSN \cite[][]{Nakamura-1987, Shibata-Nakamura-1995, Baumgarte-Shapiro-1998} formalism. We adopt the ``moving puncture'' method \citep{Zlochower-2005, Campanelli-2006, vanMeter-2006}, and our initial metric data are of the Bowen-York type \cite[][]{Bowen-1980}, conditioned to satisfy the constraint equations using the \texttt{TwoPunctures} thorn \cite[][]{Ansorg-2004}.
The GRMHD equations were solved with the \texttt{IllinoisGRMHD} code \cite[][]{Noble-2006, Etienne-2015}. 

The main structure of the paper is as follows. In Sec. \ref{sec:numerical}, we give a brief description of the numerical methods adopted in our simulations. The initial configuration of our binary evolutions are described in Sec. \ref{sec:initdata}. In Sec. \ref{sec:results}, we present results from all models considered: the dynamics of the plasma surrounding the BHs across evolution (\ref{sec:results_gasdyn}), the magnetic field enhancement and the formation of magnetically-dominated regions (\ref{sec:results_Bfield}), the mass accretion rate both during the orbital evolution and in the postmerger (\ref{sec:results_Mdot}), the development of strong Poynting flux emission (\ref{sec:results_poyn}).
\section{Numerical Methods}\label{sec:numerical}
We consider three families of simulations, each defined by BHs spin parameter $a_{1,2} = (0, \ 0.3, \ 0.6)$; for each family, we run three simulations characterized by different degrees of initial ``magnetization'', i.e., different values of initial magnetic-to-gas pressure ratio $\beta^{-1}_0$. All runs consider black holes immersed in an adiabatic gas with initial uniform density and pressure. We take the gas to be either unmagnetized (B0 models) or moderately magnetized (B1 and B2 models), set with initial uniform magnetic field aligned with the total angular momentum of the system.

In this section we give a brief overview of the mathematical setup used for producing the simulations discussed in the following. For more detailed discussion on the numerical framework adopted for evolving BH binaries in general relativity see, e.g., Refs. \cite{baumgarte-shapiro-2010, rezzolla-zanotti-2013}.
\subsection{Evolution of Gravitational Fields}\label{sec:evoGmunu}
All the equations presented below are in geometrized units ($G = c = 1$). In these units, Einstein's field equations of general relativity are
\begin{equation}\label{eq:Einstein}
 G^{\mu\nu} = 8\pi T^{\mu\nu}
\end{equation}
where $G^{\mu\nu}$ is the Einstein tensor and $T^{\mu\nu}$ the total stress-energy tensor. For a magnetized fluid, the stress-energy tensor is the sum of matter and EM components:
\begin{equation}
    T^{\mu \nu}=T_{\text {matter}}^{\mu \nu}+T_{\mathrm{EM}}^{\mu \nu}\,,
\end{equation}
\begin{equation}
    T_{\text {matter }}^{\mu \nu}=\rho h u^{\mu} u^{\nu}+p_{\mathrm{fluid}} \ g^{\mu \nu}\,,
\end{equation}
\begin{equation}
    T_{\mathrm{EM}}^{\mu \nu}=b^{2}\left(u^{\mu} u^{\nu}+\frac{1}{2} g^{\mu \nu}\right)-b^{\mu} b^{\nu}\,,
\end{equation}
where $g^{\mu\nu}$ is the metric tensor, $\rho$ is the rest-mass density, $u^{\mu}$ is the 4-velocity of the fluid, $h$ is the specific enthalpy, $p_{\mathrm{fluid}}$ is the fluid pressure, and $b^{\mu}$ is the magnetic 4-vector. The space-time metric in standard 3+1 form is
\begin{equation}\label{eq:3+1lineel}
 d s^{2}=-\alpha^{2} d t^{2}+\gamma_{i j}\left(d x^{i}+\beta^{i} d t\right)\left(d x^{j}+\beta^{j} d t\right)
\end{equation}
where $\alpha$ is the lapse function, $\beta^i$ the $i$th component of the shift vector and $\gamma_{ij}$ is the spatial metric. The extrinsic curvature $K_{ij}$ is given by
\begin{equation}\label{eq:Kij}
 \left(\partial_{t}-\mathcal{L}_{\beta}\right) \gamma_{i j}=-2 \alpha K_{i j}
\end{equation}
with $\mathcal{L}_{\beta}$ denoting the Lie derivative with respect to $\beta$.
We evolve the metric variables $(\gamma_{ij}, K_{ij})$ using the BSSN formulation (BSSN evolution and constraint equations are summarized in Refs. \cite{Shibata-Nakamura-1995, Baumgarte-Shapiro-1998}).

Our metric evolution equations do not include matter source terms, since for all the simulations considered in this work we assume that the total mass of the fluid is negligible with respect to the mass of the two BHs, $M_{\mathrm{fluid}} \ll M_{\mathrm{BHs}}$ (i.e., we evolve the Einstein equations in vacuum).
We adopt the ``1+log'' slicing condition for the lapse and a ``hyperbolic gamma-driving'' condition for the shift \cite{vanMeter-2006}. 

In general, nonradiative GRMHD simulations are scale free. Thus, we will use length and time units that scale with the total mass of the system $M$. All our simulations are evolved setting $c=G=M=1$. The unitary values of mass, length and time in code units correspond to
    \begin{equation}\label{eq:A-mass}
       \hat{m} = M = 2\cdot10^{39}M_6 \ \mathrm{g}
    \end{equation}
    \begin{equation}\label{eq:A-length}
        \hat{l} = \frac{G}{c^2}M = 1.48\cdot 10^{11}M_6 \ \mathrm{cm}
    \end{equation}
    \begin{equation}\label{eq:A-time}
        \hat{t} = \frac{G}{c^3}M = 4.94M_6 \ \mathrm{s}
    \end{equation}
where $M_6\equiv M/10^6\mathrm{M}_{\odot}$.
Since we assume $M_{\mathrm{fluid}} \ll M_{\mathrm{BHs}}$, the fluid contribution to Eq. \eqref{eq:Einstein} can be ignored, and we can set $T^{\mu\nu} \approx 0$. This means that we are free to rescale an appropriate set of the fluid field variables independently of the geometric scaling that arises by the condition $M=1$.

\begin{table}
 \caption{BBH initial data parameters and derived quantities in code units of the GRMHD runs: initial puncture separation $d_0$ and linear momentum components $p_x$ \& $p_y$, dimensionless spin parameter $a$ of each BH, merger time $t_{\mathrm{Merger}}$ and initial ratio of magnetic-to-fluid pressure $\beta^{-1}_0$.}\label{tab:initdata}
 \begin{tabular}{lccccccc}
\hline\hline \\[-1.6ex]
  Run & $d_0$ & $p_{x}$ & $p_{y}$ &  $a_{1,2}$ & $t_{\mathrm{Merger}}$  &  $\beta^{-1}_0$  \vspace{0.2pt}\\
\hline
  B0S0 &   &  &  &  &   &    0 \\
\cmidrule(lr) {1-1}\cmidrule(lr) {7-7}
  B1S0 &  12.038 & 5.26$\cdot 10^{-4}$ & 0.085 &  0.0  &  1834  &   0.025 \\
\cmidrule(lr) {1-1}\cmidrule(lr) {7-7}
  B2S0 &   &  &  &   &  &  0.31 \\
\hline
  B0S3 &   &  &  &   &    &  0 \\
\cmidrule(lr) {1-1}\cmidrule(lr) {7-7}
  B1S3 &  12.162 & 4.78$\cdot 10^{-4}$ & 0.083 &  0.3  &   2198 & 0.025 \\
\cmidrule(lr) {1-1}\cmidrule(lr) {7-7}
  B2S3 &   &  &  &   &   &   0.31 \\
\hline
  B0S6 &   &  &  &   &    &  0 \\
\cmidrule(lr) {1-1}\cmidrule(lr) {7-7}
  B1S6 &  12.162 & 4.62$\cdot 10^{-4}$ & 0.082 &  0.6  &   2540 & 0.025 \\
\cmidrule(lr) {1-1}\cmidrule(lr) {7-7}
  B2S6 &   &  &  &   &   &   0.31 \\
\hline\hline
 \end{tabular}
\end{table}
\subsection{Evolution of magnetohydrodynamic fields}
The GRMHD equations and constraint equations are derived from the following:
\begin{enumerate}
 \item the conservation of baryon number
 \begin{equation}\label{eq:baryon-cons}
 \nabla_{\mu}(\rho u^{\mu}) = 0\,,
  \end{equation}
\item the conservation of energy momentum
\begin{equation}\label{eq:enmom-cons}
 \nabla_{\mu}T^{\mu\nu}=0\,,
\end{equation}
where $T^{\mu\nu}= T^{\mu\nu}_{\mathrm{matter}} + T^{\mu\nu}_{\mathrm{EM}}$, and
\item the homogenous Maxwell's equations
\begin{equation}\label{eq:hom-maxwell}
 \nabla_{\nu} F^{* \mu \nu}=\frac{1}{\sqrt{-g}} \partial_{\nu}\left(\sqrt{-g} F^{* \mu \nu}\right)=0\,,
\end{equation}
where $F^{\mu\nu}$ is the Faraday tensor, $F^{*\mu\nu}$ is its dual, and $g$ is the determinant of $g_{\mu\nu}$.
\end{enumerate}
The \texttt{IllinoisGRMHD} code \cite{Etienne-2015} evolves a set of conservative MHD fields $\boldsymbol{C} \equiv \left\{\rho_{*}, \tilde{\tau}, \tilde{S}^i, \tilde{B}^i\right\}$ solving the coupled Einstein-Maxwell equations. It  assumes a perfect fluid stress-energy tensor for the matter and infinite conductivity (ideal MHD limit). The vectors of the ``conservative variables'' $\boldsymbol{C}$ depend directly on the ``primitive variables'' $\boldsymbol{P} \equiv \left\{\rho, p_{\mathrm{fluid}}, v^{i}, B^{i}\right\}$ where $\rho$ is the rest-mass density, $p_{\mathrm{fluid}}$ is the fluid pressure, $v^i\equiv u^i/u^0$ are the components of the fluid three-velocity and $B^i$ are the spatial components of magnetic field $B^{\mu}$ measured by Eulerian observers.

To satisfy the divergence-free nature of the magnetic field, the \texttt{IllinoisGRMHD} code evolves the magnetic four-vector potential $\mathcal{A}_{\mu}$ instead of the magnetic fields directly (see \cite{Etienne-2015}), so that
\begin{equation}
    \begin{array}{l}
\mathcal{A}_{\mu}=\Phi n_{\mu}+A_{\mu}, \\
\tilde{B}^{i}=\tilde{\epsilon}^{i j k} \partial_{j} A_{k}
\end{array}
\end{equation}
where $A_{\mu}$ is purely spatial ($A_{\mu} n^{\mu}=0$) and $\Phi$ is the EM scalar potential. The standard permutation symbol $\tilde{\epsilon}^{i j k} $ is equal to 1 (-1) if $ijk$ are an even (odd) permutation of 123, and 0 if one or more indices are identical.

We apply the so-called ``outflow'' boundary conditions to the hydrodynamic variables ($\rho_0, p_{\mathrm{fluid}}, v^i$) and a linear extrapolation to $A_{\mu}$ \cite{Etienne-2015}.
\subsection{Magnetized Accretion Flows}
At present, our understanding of accretion flow properties around merging MBHBs is uncertain. On very small scales (such as the ones we consider), it is not possible to uniquely define initial conditions for the gas in the vicinity of merging binaries. Following \citeAliasTwo{Giacomazzo-2012}{Gi12} and \citeAliasTwo{Kelly-2017}{Ke17}, we choose to evolve our models in a simple environment consisting in a homogenous, ideal gas with initial uniform rest-mass density $\rho_0$, which has an initially uniform magnetic field (aligned with the orbital angular momentum) and fills the entire computational domain.

An ideal gas with adiabatic index $\gamma$ has a pressure
\begin{equation}
   p_{\mathrm{fluid}} = \rho\epsilon(\gamma-1)
\end{equation}
(where $\rho$ is the rest-mass density, $\epsilon$ is the specific internal energy, and $\gamma$ is the adiabatic index), and a specific enthalpy
\begin{equation}
   h=(1+\epsilon)+\frac{p_{\mathrm{fluid}}}{\rho}=1+\gamma \epsilon
\end{equation}
We also choose our gas to obey the polytropic equation of state
\begin{equation}
   p_{\mathrm{fluid}}=\kappa\rho^{\Gamma},
\end{equation}
with a polytropic index $\Gamma=4/3$ and a polytropic constant $\kappa$ to be assigned. We assume that the adiabatic index of the fluid $\gamma$ is coincident with the polytropic index $\Gamma$; hence, the specific internal energy of the gas can be expressed as
\begin{equation}
   \epsilon=\frac{\kappa \rho^{\Gamma-1}}{\Gamma-1}
\end{equation}
The speed of sound in the gas is

\begin{equation}\label{eq:cs}
   \begin{split}
 c_{\mathrm{s}} = \ & \sqrt{\frac{1}{h}\left(\frac{\partial p_{\mathrm{fluid}}}{\partial \rho}+\frac{p_{\mathrm{fluid}}}{\rho^{2}} \frac{\partial p_{\mathrm{fluid}}}{\partial \epsilon}\right)}\\
 = \ & \sqrt{\frac{(\Gamma-1)(\epsilon+p_{\mathrm{fluid}} / \rho)}{h}}\\
 = \ & \sqrt{\frac{\Gamma p_{\mathrm{fluid}}}{h\rho}}
   \end{split}
\end{equation}
We express the magnetic field with the magnetic four-vector $b^{\mu}$ \cite[see, e.g., Ref. ][]{Duez-2005}:
\begin{equation}
   b^{\mu}=\frac{1}{\sqrt{4 \pi} \alpha}\left(u_{m} B^{m}, \frac{B^{i}+\left(u_{m} B^{m}\right) u^{i}}{u^{0}}\right)
\end{equation}
where repeated latin indices indicate sums over spatial components only.

The relativistic Alfv\'{e}n velocity of a magnetized plasma \cite[][]{Gedalin-1993} is defined  as 
\begin{equation}\label{vAlf-def}
    \begin{aligned}
v_{\mathrm{Alf}} &=\sqrt{\frac{b^{2}}{\rho(1+\epsilon)+p_{\mathrm{fluid}}+b^{2}}}=\sqrt{\frac{b^{2}}{\rho(1+\Gamma \epsilon)+b^{2}}} \\
&=\sqrt{\frac{b^{2}}{\rho+4 p_{\mathrm{fluid}}+b^{2}}}
\end{aligned}
\end{equation}
where the second line holds for polytropic fluids with $\Gamma = 4/3$.
\subsection{Diagnostics}
To explore the effects of spin and magnetic field strength on the dynamics of the accreting gas we track the evolution of the following quantities:

\begin{itemize}
    \item rest-mass density $\rho$ (normalized to its initial value $\rho_0$);
    \item Newtonian Mach number $\mathscr{M} \equiv \mathrm{v}/c_s$, where $\mathrm{v}$ is defined as the velocity magnitude of the fluid on the orbital plane 
    \begin{equation}\label{eq:gasspeed}
        \mathrm{v}=(\mathrm{v}_x^2+\mathrm{v}_y^2)^{1/2}
    \end{equation}
    and  $c_s$ is the speed of sound in the medium (Eq. \eqref{eq:cs});
    \item Newtonian angular velocity $\Omega_{\mathrm{fluid}}$ of the fluid about the orbital axis, defined as 
\begin{equation}\label{eq:def-omegaF}
    \Omega_{\mathrm{fluid}} = \frac{xv_y-yv_x}{(x^2+y^2)}.
\end{equation}
The quantity $\Omega_{\mathrm{fluid}}$ has the dimension of $\hat{t}^{-1}$.
\end{itemize}
Given a test particle orbiting a Kerr BH of mass $M$ and spin parameter $a$, the coordinate angular frequency of a circular orbit (for those values of $r$ for which circular orbits exist) is \cite[see, e.g., Ref. ][]{Bardeen-1972}
\begin{equation}
    \Omega_{\mathrm{circ}}^{\pm} = \  \pm\frac{M^{1/2}}{r^{3/2}\mp aM^{1/2}}
\end{equation}
where $r$ is the areal radius in Kerr-Boyer-Lindquist (KBL) coordinates and the sign +(-) refers to corotating (counterrotating) orbits. We define a \textit{circularity parameter} $\omega$ as
\begin{equation}\label{eq:rotparameter}
    \omega \equiv \Omega_{\mathrm{fluid}}/\Omega_{\mathrm{circ}}^{+}
\end{equation}
Following the evolution of these diagnostics allows us to better interpret the results of each simulation (e.g., monitoring $\mathscr{M}$ and $\omega$ will help us studying the degree of rotation induced on the gas by the inspiralling BHs).

In Sec. \ref{sec:results_poyn} we explore how the evolution of the magnetic fields affects the possible emission of EM signals. As was pointed out in Refs. \cite{Palenzuela-2010, Moesta-2010} (for electrovacuum) and \cite{Palenzuela-2010b, Moesta-2012} (for force-free plasma), the later inspiral and merger of massive BH binaries immersed in a magnetized gas may be connected with an EM counterpart in the form of a jet, which could be potentially visible at large distances. In this work we study this strong and collimated electromagnetic emission looking primarily at the Poynting vector \cite[][]{Giacomazzo-2012, Kelly-2017}. It is calculated as
\begin{equation}\label{eq:poynting-def}
    S^{i} \equiv \alpha T_{\mathrm{EM}, 0}^{i}=\alpha\left(b^{2} u^{i} u_{0}+\frac{1}{2} b^{2} g_{0}^{i}-b^{i} b_{0}\right)
\end{equation}

\section{Initial Data}\label{sec:initdata}
We reexamine the setup of \citeAliasTwo{Kelly-2017}{Ke17}, performing simulations of MMAFs onto binaries of equal-mass BHs.
The individual mass of each BH in code units is $M_{\mathrm{BH}}=M/2 =0.5$, where $M=1$ is the total mass of the system. Our binaries are immersed in an initially uniform, radiation-dominated polytropic fluid ($p_0 = \kappa \rho_0^{\Gamma}$, with $\rho_0 = 1$, \ $\kappa = 0.2, \ \Gamma = 4/3$). To capture the effect of the individual BHs spins on the accretion flows, we evolve binaries of spinning BHs with parallel spins aligned with the orbital axis, and adimensional spin parameters $a_1=a_2 = (0, \ 0.3, \ 0.6)$.

We adopt a cubical domain given by $[-1024M, 1024M]^3$ and employ AMR with $N=11$ levels of refinement. The coarsest resolution is $\Delta x_c = 64M/3$, and the finest one is $\Delta x_f = \Delta x_c \cdot 2^{1-N} = M/48$. All our simulations could be easily rescaled to consider systems of binary black holes with a total mass $M = 2\times 10^6 \ \mathrm{M}_{\odot}$, and immersed in a gas with uniform initial rest-mass density $\rho_0 = 10^{-11}$ g cm$^{-3}$. These values are consistent with the approximation $T^{\mu\nu} \approx 0$, since they yield $M_{\mathrm{fluid}}/M_{\mathrm{BH}} \sim 10^{-7}$ (see Sec. \ref{sec:evoGmunu}).

The BHs rotate around each other starting on quasicircular orbits at an initial separation $d_0 \simeq 12 M$.
We decided to set our initial separations to $12 M$ on the basis of previous results of \citeAliasTwo{Kelly-2017}{Ke17}. The simulations by \citeAliasTwo{Kelly-2017}{Ke17} consider initial separations covering values between $6.6 M$ and $16.3 M$. It was found that features in the evolution of the Poynting luminosity are well resolved for initial separations $\gtrsim 11.5 M$.
More specifically, simulations by \citeAliasTwo{Kelly-2017}{Ke17} of binaries with separations of $11.5 M$, $14.4 M$, and $16.3 M$ show the same qualitative behavior, thus we chose to evolve our binaries starting from an initial separation of $12 M$. This choice allows for the evolution of different configurations up to and beyond merger.

Our quasicircular initial data are obtained from larger-scale PN evolutions (the PN equations are evolved from a larger separation, $\sim40 M$, to the distance we begin our full GR runs with). In Table \ref{tab:initdata} we give the initial data for the nine configurations presented in this paper.

\subsection{Initial plasma configuration}
We evolve MBHBs immersed in a hot plasma, which is threaded by an initially uniform magnetic field parallel to the binary angular momentum, i.e. $B^i=(0,0,B^z)$. The magnetic field is assumed to be anchored to a distant circumbinary disk located outside the computational domain. This initial configuration of the magnetic field is analogous to that implemented in previous works (e.g., Refs. \cite{Palenzuela-2010, Palenzuela-2010b, Moesta-2012}, \citeAliasTwo{Giacomazzo-2012}{Gi12}, \citeAliasTwo{Kelly-2017}{Ke17}).
While simplistic, our choice of the initial plasma configuration is sufficiently clear to aid in pinpointing the effects of subtle physical processes (e.g., the spins) on the accretion flows.
We set three different initial plasma configurations, which are chosen so that
\begin{equation}\label{eq:beta0}
    \beta^{-1}_0 \equiv \frac{p_{\mathrm{mag}}}{p_{\mathrm{fluid}}} = \left\{\begin{array}{lr}
         0 \ \ \ \ \ \ \ \ \ \  &\text{(B0 runs)}  \\
         0.025 \ \ \ \ \  &\text{(B1 runs)} \\
         0.31 \ \ \ \ \ \ \  &\text{(B2 runs)}
    \end{array}\right.
\end{equation}
or, equivalently,
\begin{equation}\label{eq:zeta}
    \zeta_0 \equiv \frac{u_{\mathrm{mag}}}{u_{\mathrm{fluid}}} = \left\{\begin{array}{lr}
         0 \ \ \ \ \ \ \ \ \ \ &\text{(B0 runs)}  \\
         0.005 \ \ \ \ \ &\text{(B1 runs)} \\
         0.063 \ \ \ \ \ &\text{(B2 runs)}
    \end{array}\right.
\end{equation}
where $\beta^{-1}$ and $\zeta$ are the adimensional magnetic-to-fluid pressure ratio and magnetic-to-fluid energy density ratio, respectively, and
\begin{displaymath}
   \begin{split}
       &u_{\mathrm{mag}}  =  p_{\mathrm{mag}}  =  \frac{B^2}{8\pi} = \frac{b^2}{2}, \\
       &u_{\mathrm{fluid}} = \rho c^2, \ \ p_{\mathrm{fluid}} = \kappa\rho^{\Gamma}.
   \end{split}
\end{displaymath}
In Table \ref{tab:initparams} we list the initial uniform GRMHD field values for the three sets of configurations B0, B1 and B2.
\begin{table}[b]
 \begin{ruledtabular}
 \begin{tabular}{lcccccc}
  Run & $\rho_0$ & $\kappa$ & $\zeta_0$ &  $\beta^{-1}_0$ & $v_{\mathrm{alf}}$ \vspace{0.2pt}\\
\hline
  B0 & 1  & 0.2 & 0.0 & 0.0 & - \\
\midrule
  B1 & 1  & 0.2 & 5e-3 &  2.5e-2 & 7.4e-2\\
\midrule
  B2 & 1  & 0.2 & 6.3e-2 & 0.31 & 0.26\\
 \end{tabular}
 \end{ruledtabular}
  \caption{Initial uniform GRMHD field values for the three sets of configurations B0, B1 and B2: rest-mass density $\rho$, polytropic constant $\kappa$, magnetic-to-gas energy density (pressure) ratio $\zeta$ ($\beta^{-1}$), Alfv\'{e}n velocity $v_{\mathrm{alf}}$. The values are in code units.}\label{tab:initparams}
\end{table}
The value of the initial fluid rest-mass density $\rho_0 = 10^{-11}$ g cm$^{-3}$, along with a specific choice of $\zeta_0$ ($\beta_0^{-1}$), uniquely fixes the corresponding physical values of the initial magnetic field strength $B_0$ and of the initial fluid temperature $T_0$.
\begin{figure*}
\begin{center} 
\includegraphics[width=5.9cm]{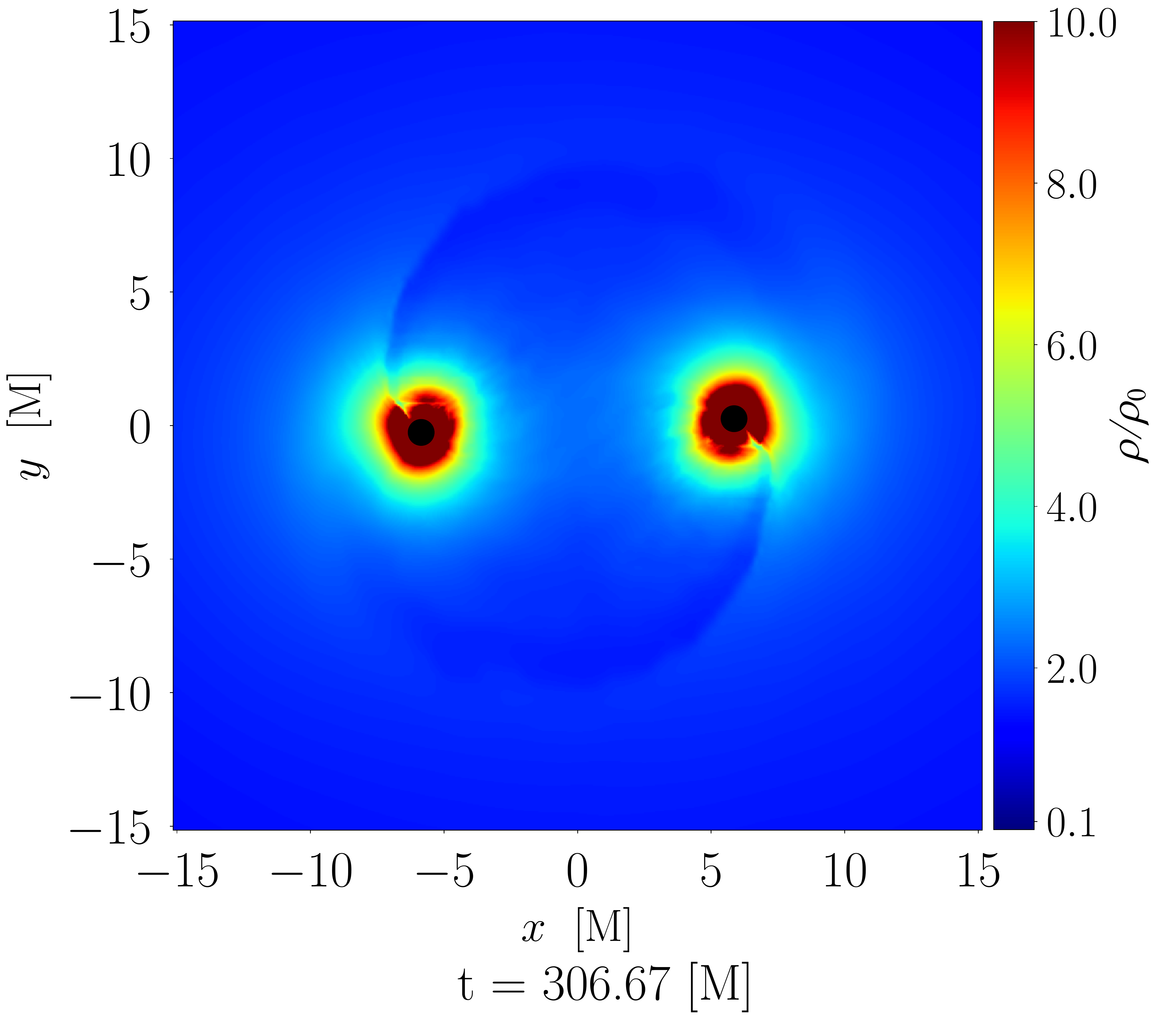}
\includegraphics[width=5.9cm]{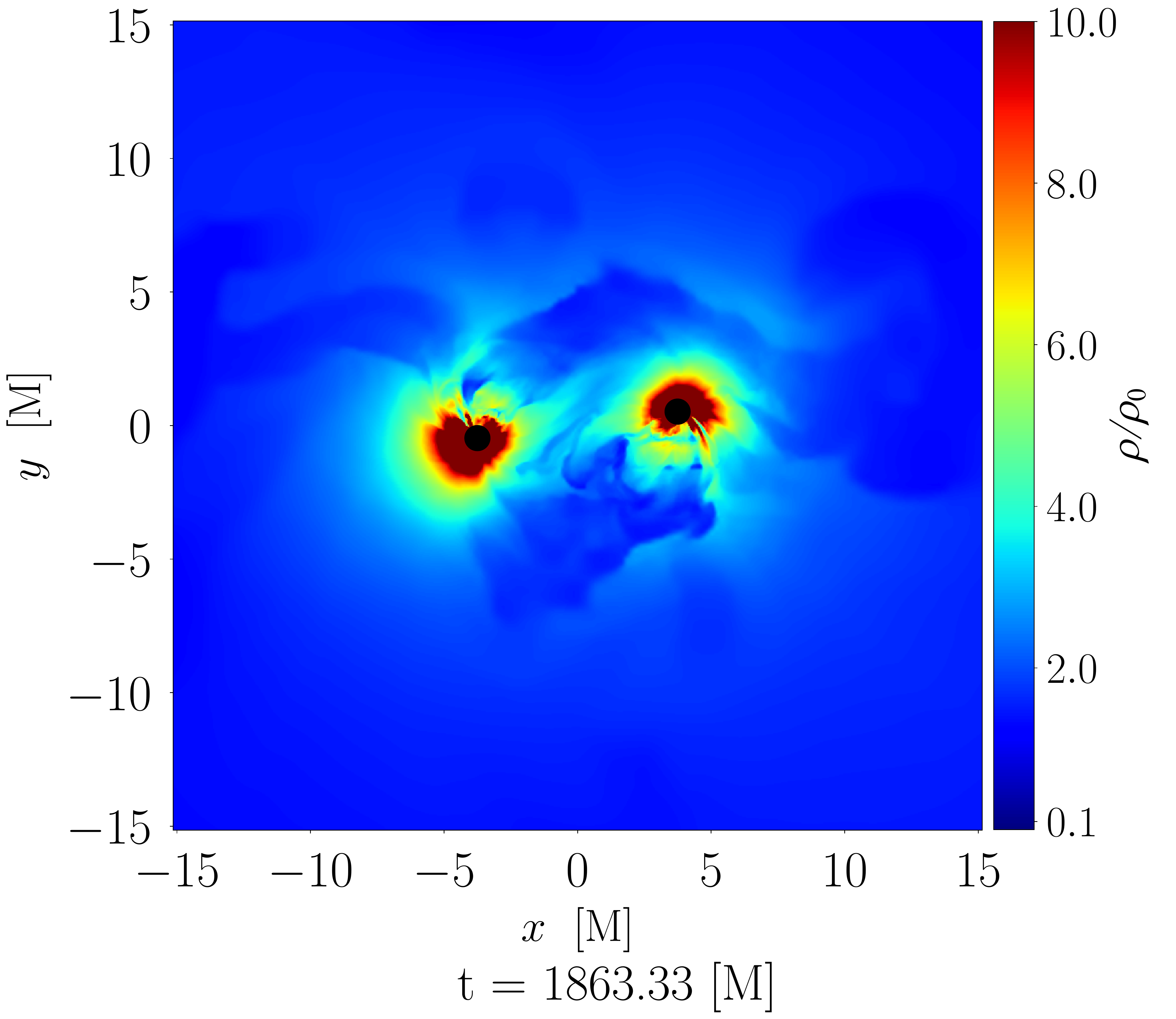}
\includegraphics[width=5.9cm]{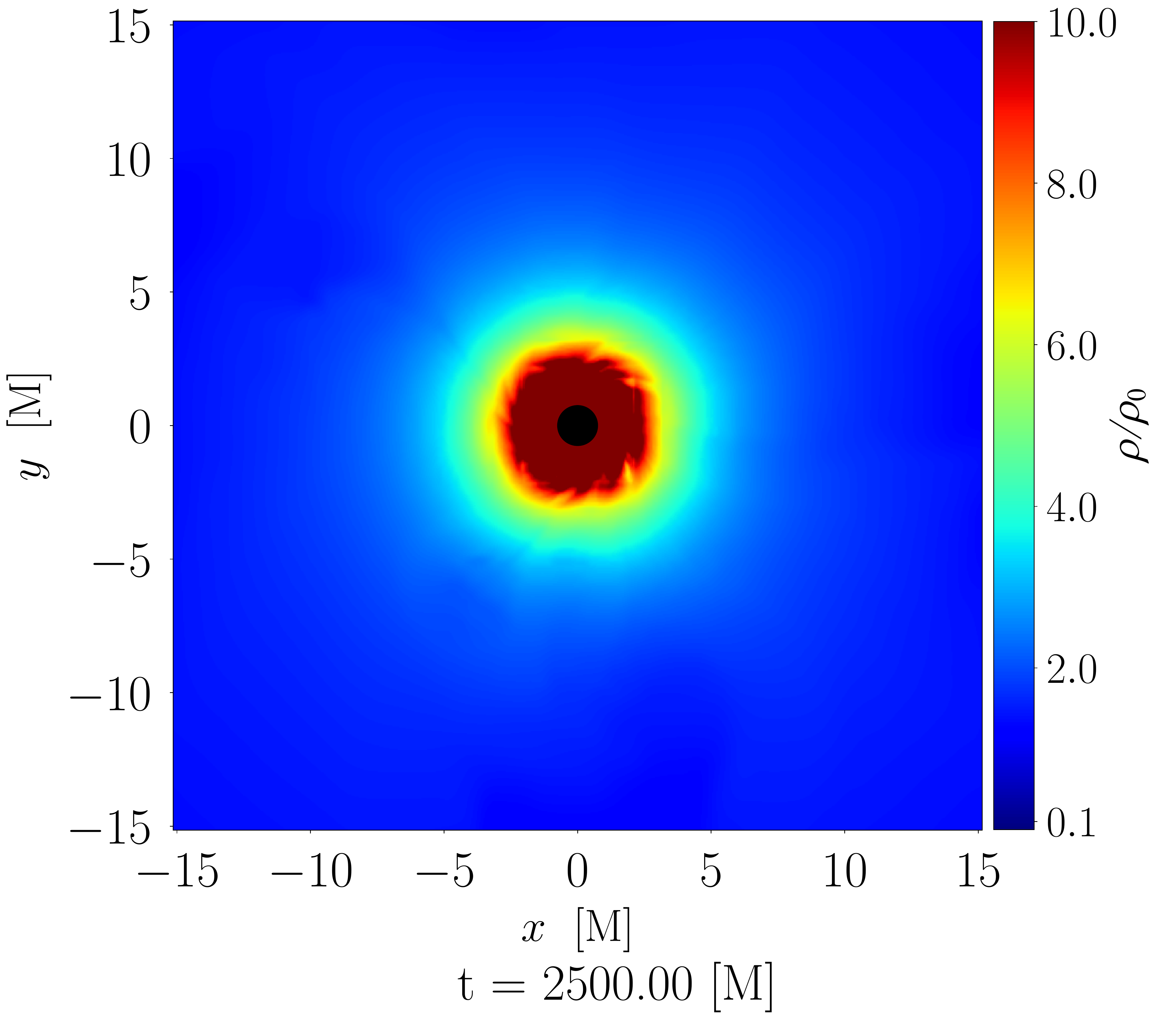}\\
\includegraphics[width=5.9cm]{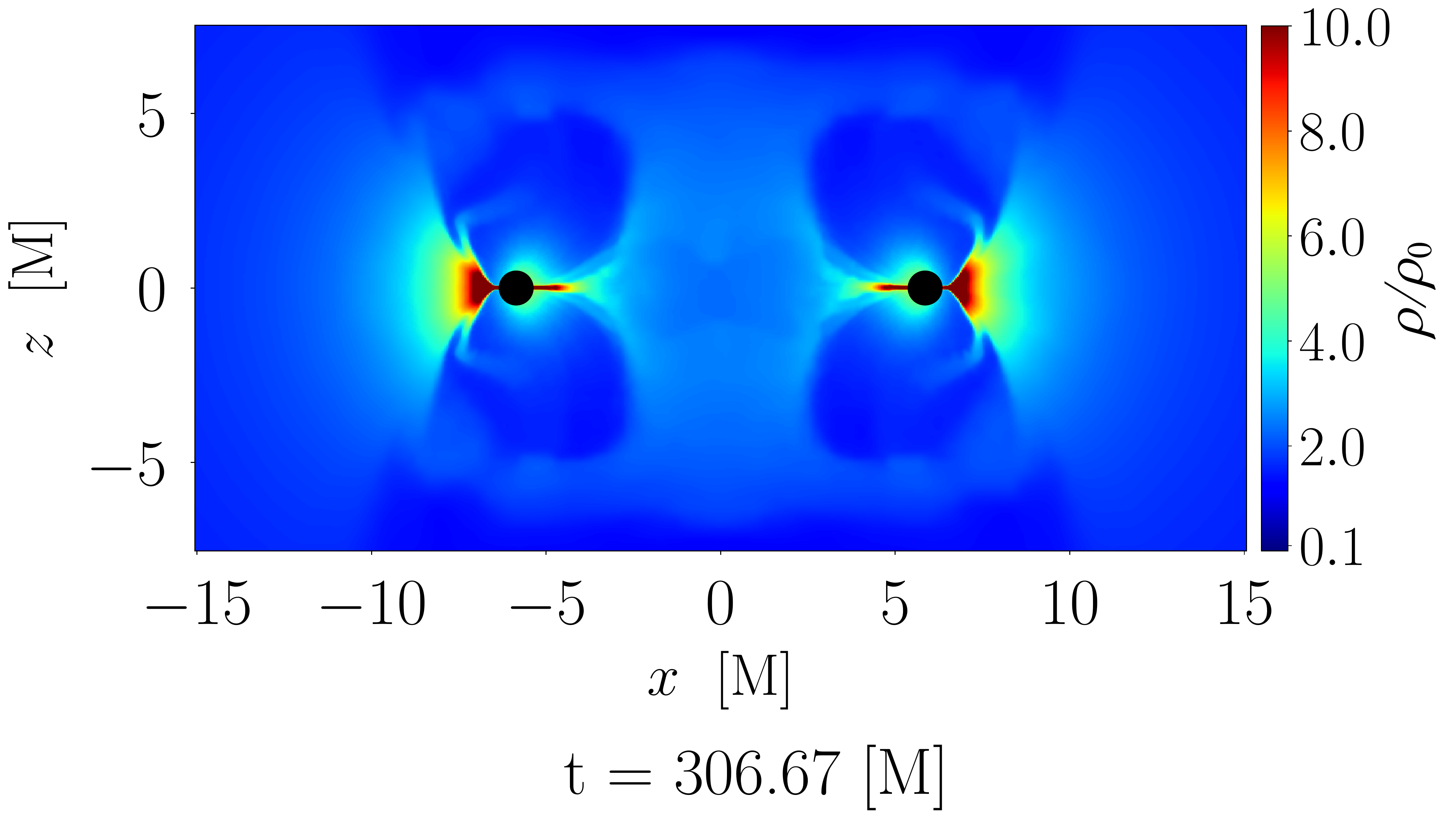}
\includegraphics[width=5.9cm]{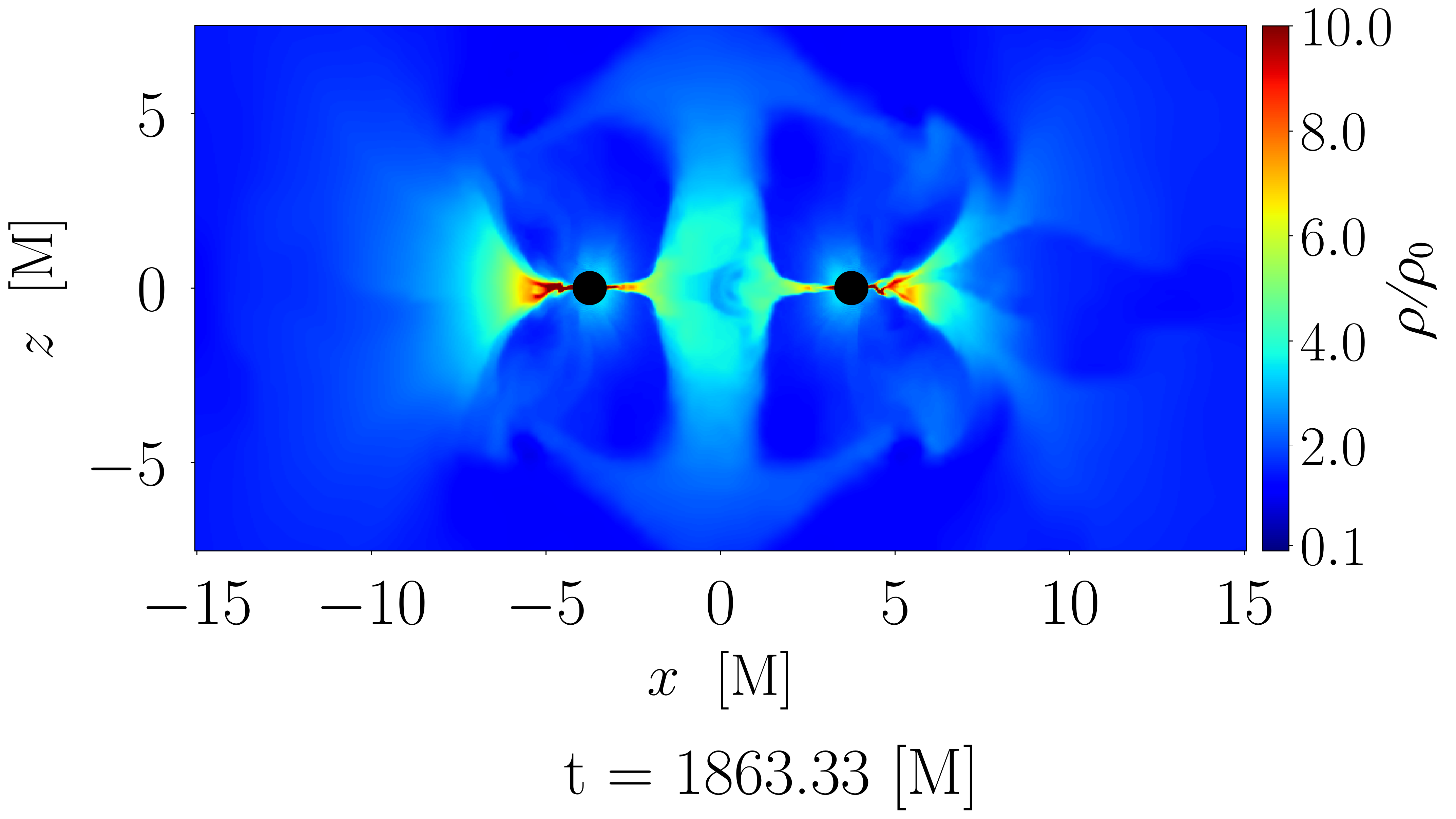}
\includegraphics[width=5.9cm]{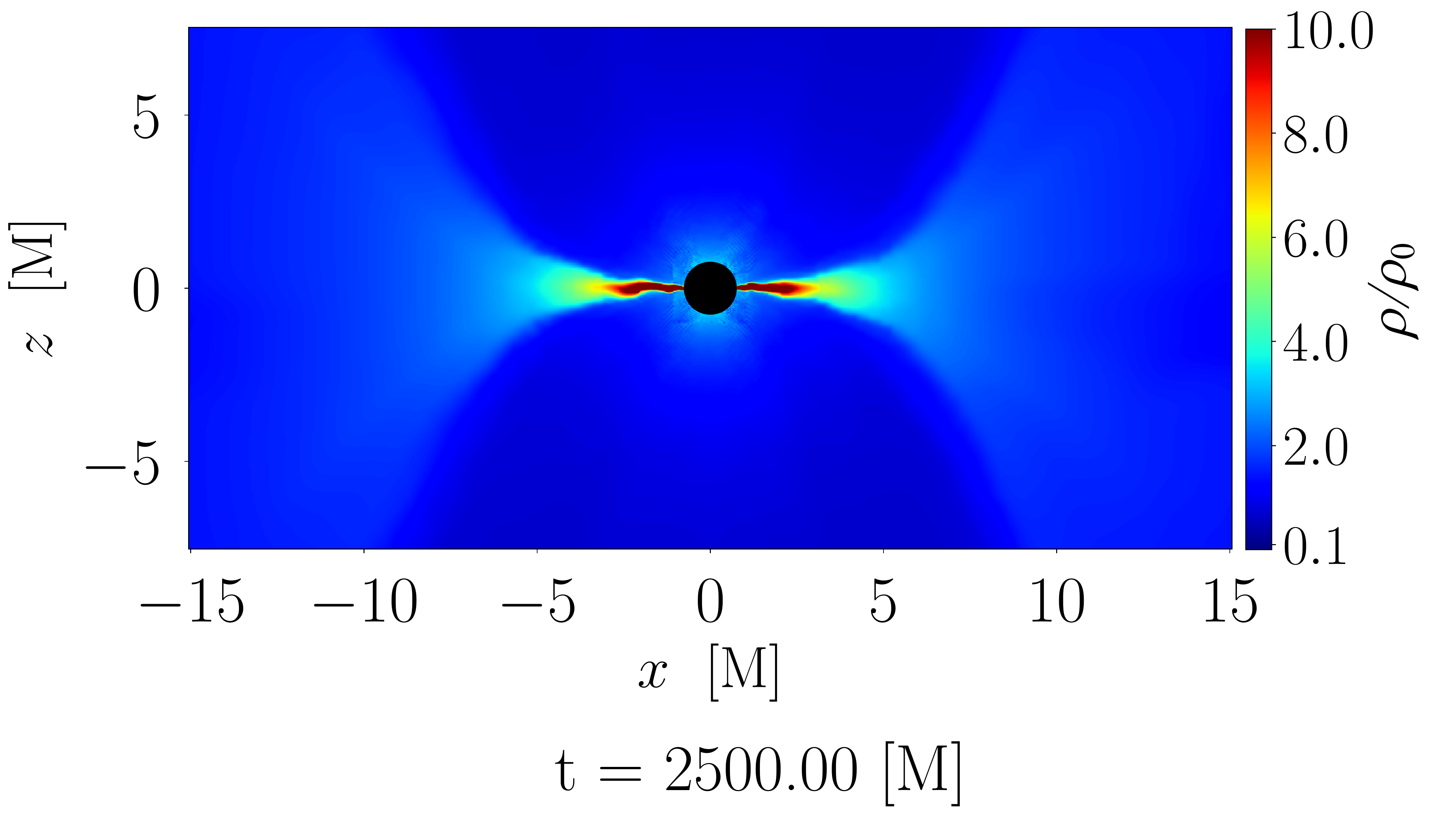}
\end{center}
\caption{Rest-mass density $\rho$ (normalized to its initial value $\rho_0$) on the $xy$ (top panels) and $xz$ planes (bottom panels) for the B2S3 configuration ($a=0.3, \beta_0^{-1}=0.31$). The snapshots were taken, respectively, after $\sim$1 orbit, after $\sim$8 orbits and at a time equal to $\sim300 \ M$ after the merger. The regions inside the BH horizons have been masked out.}\label{fig:rhoevolution-B3}
\end{figure*}
\begin{figure*}
\begin{center} 
\includegraphics[width=5.9cm]{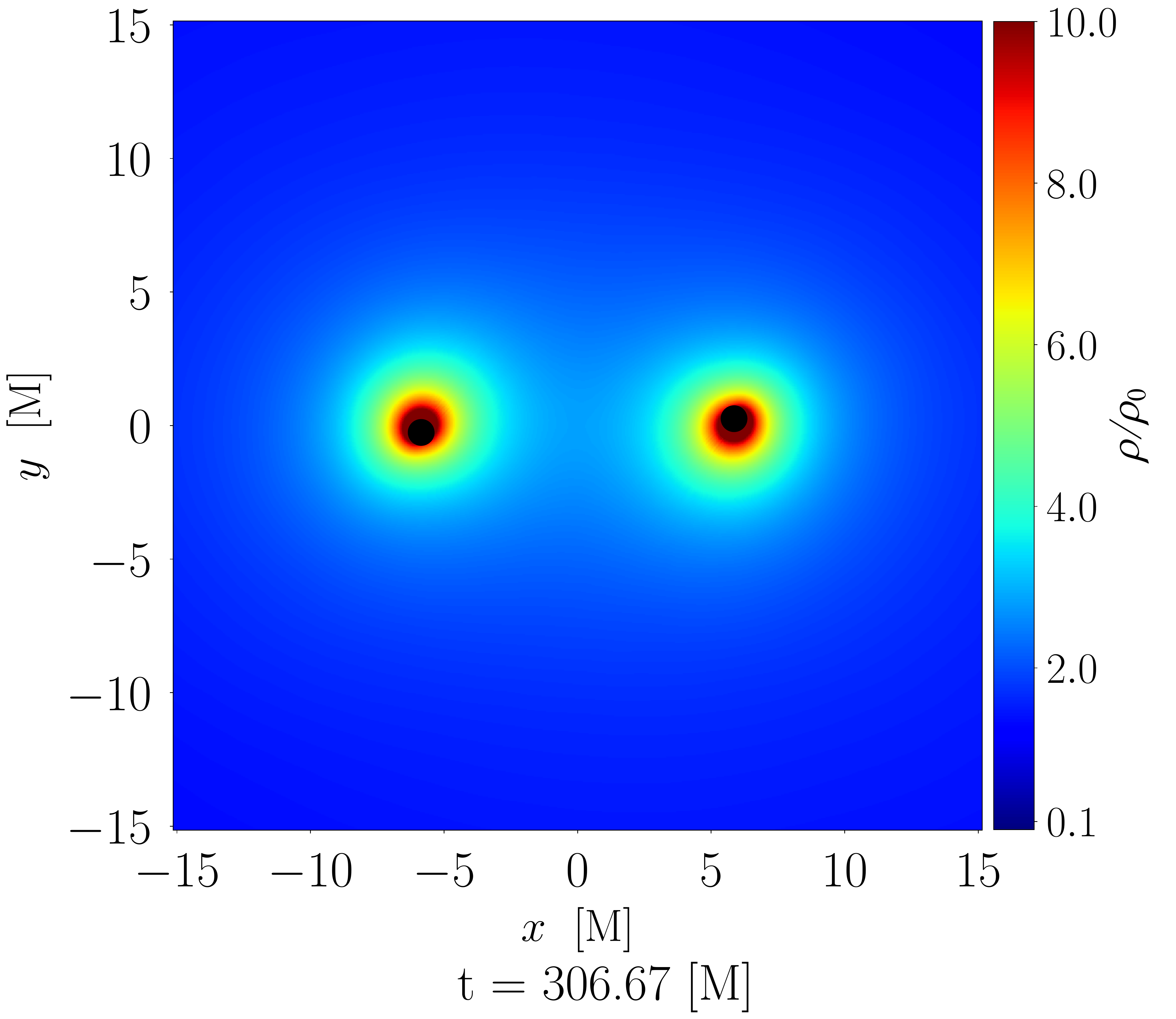}
\includegraphics[width=5.9cm]{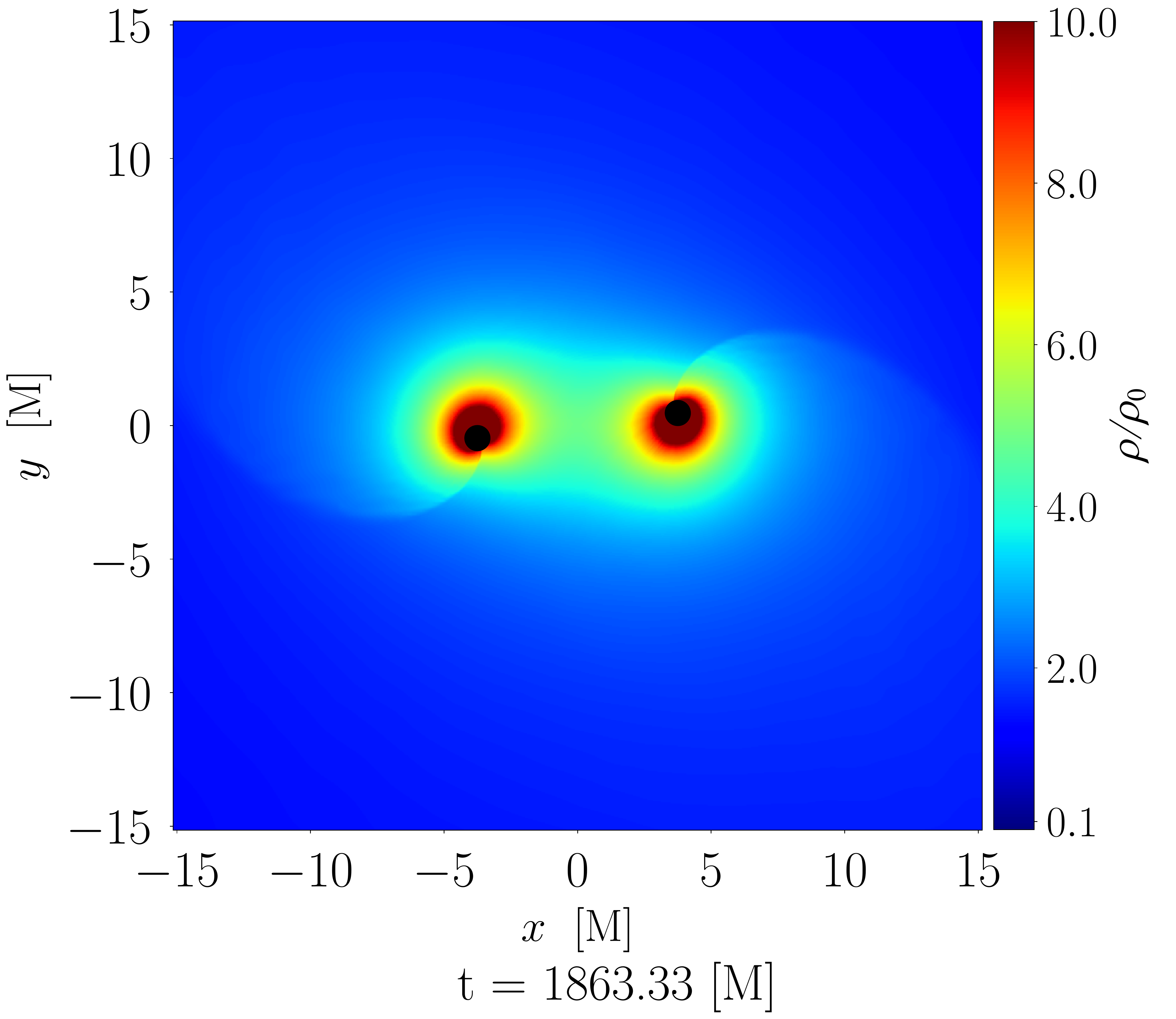}
\includegraphics[width=5.9cm]{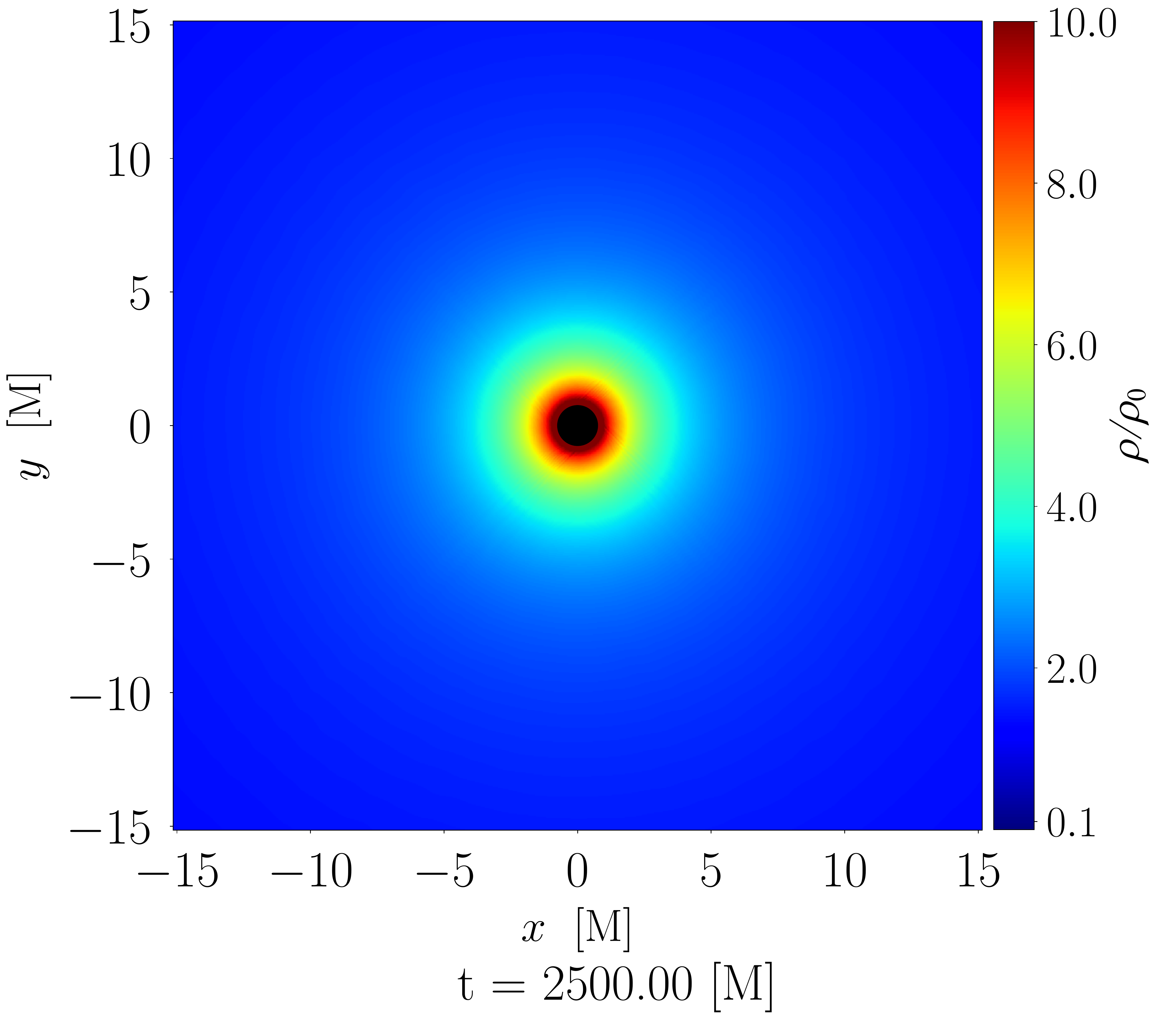}\\
\includegraphics[width=5.9cm]{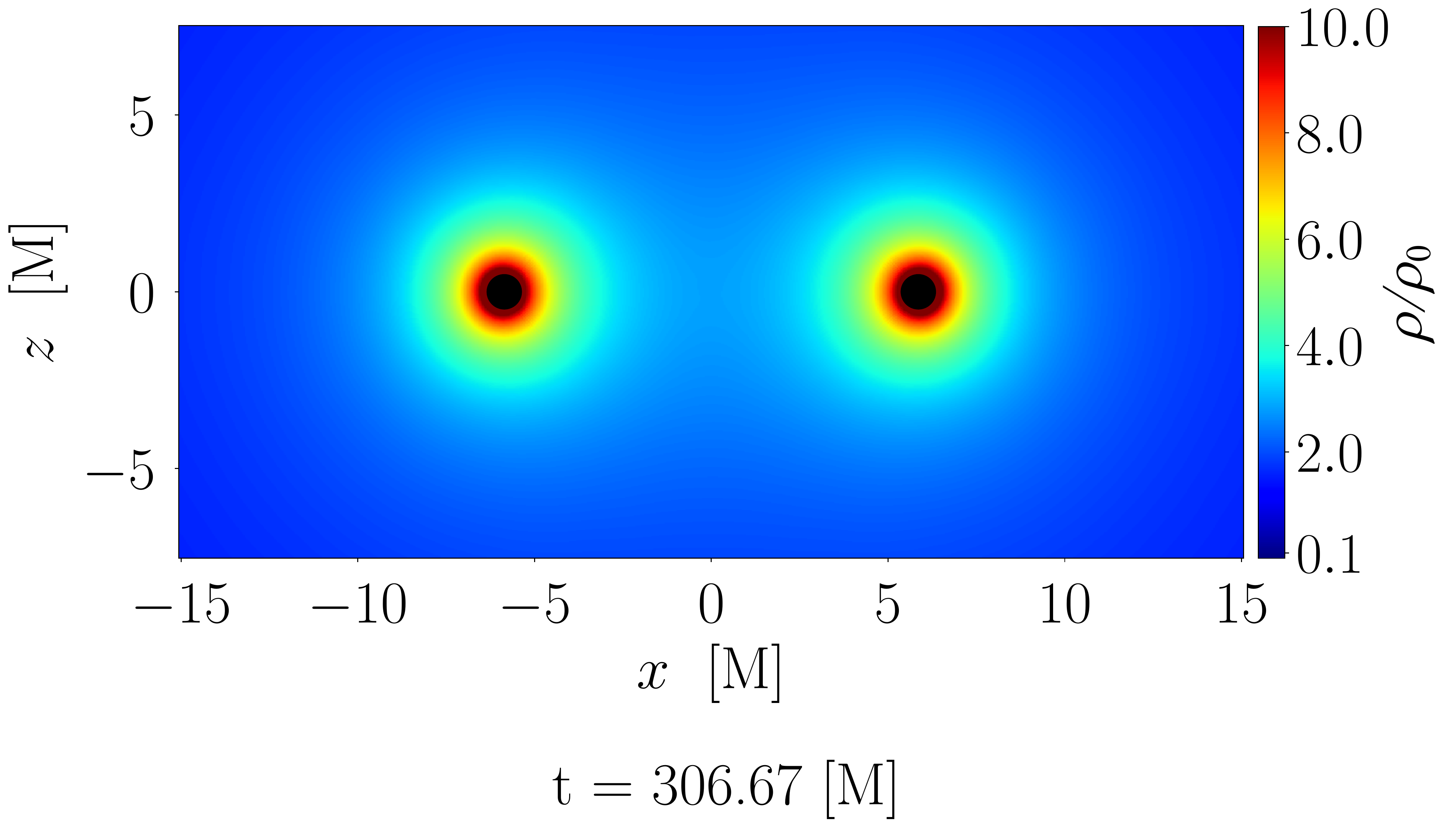}
\includegraphics[width=5.9cm]{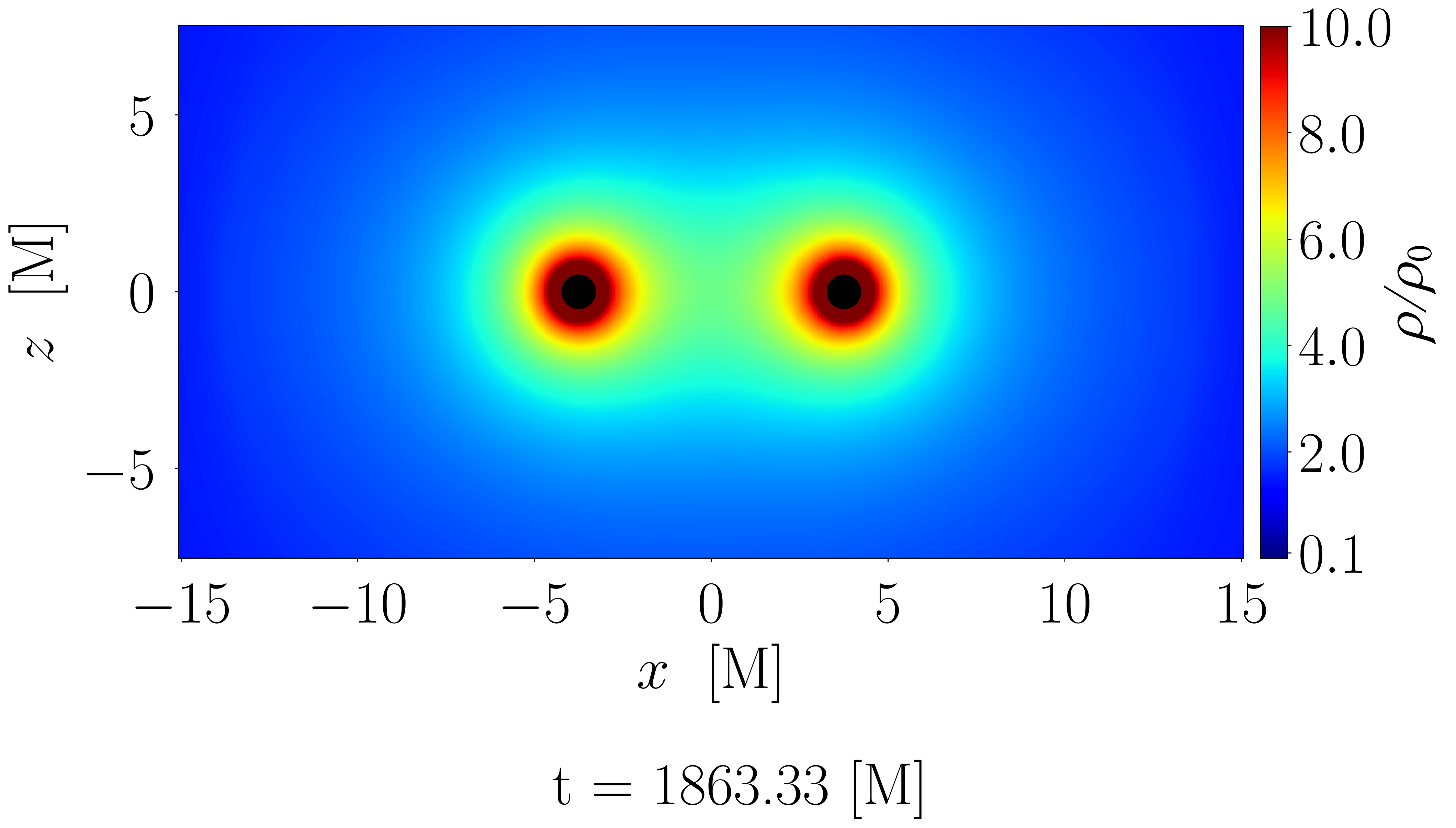}
\includegraphics[width=5.9cm]{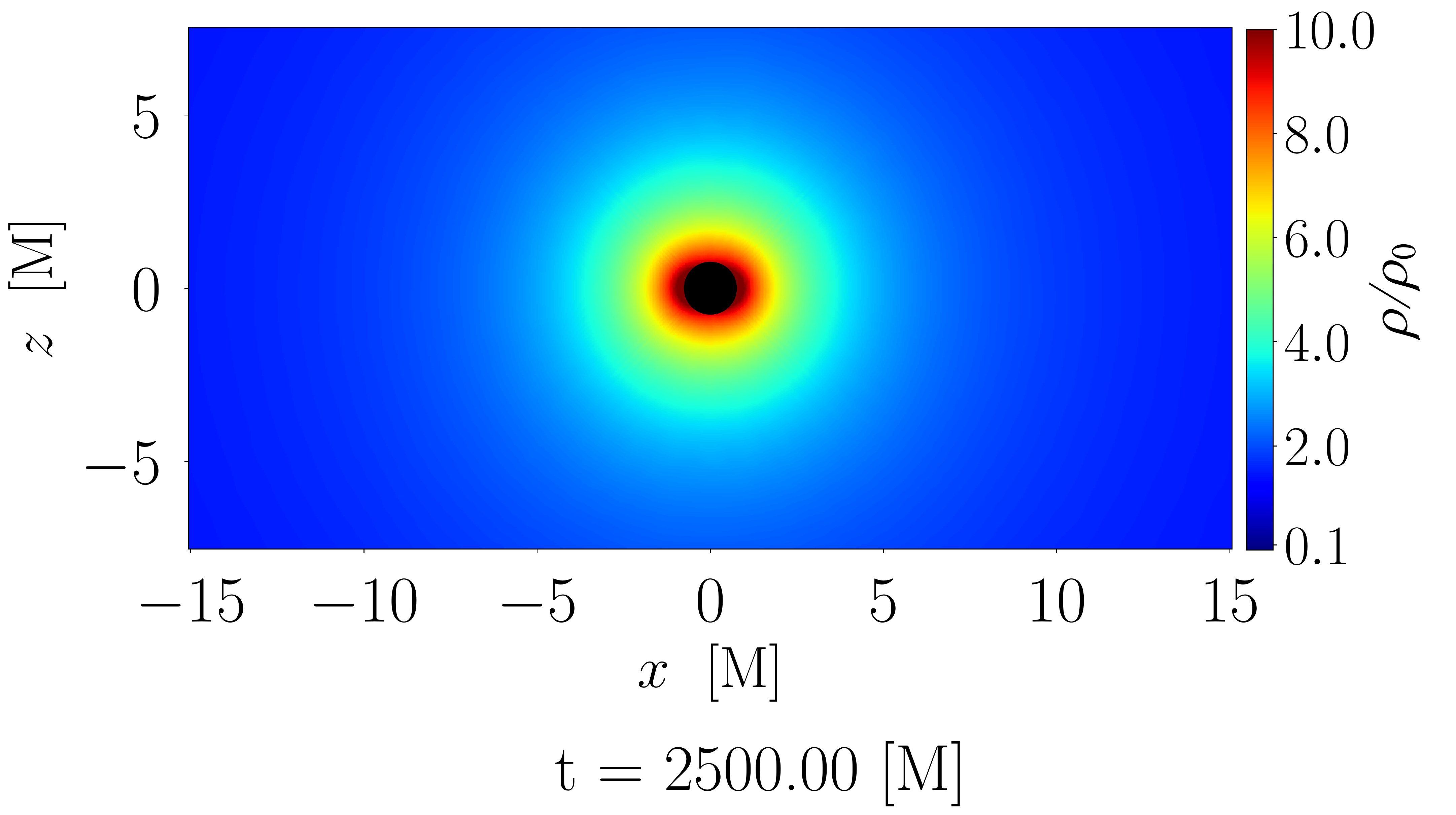}
\end{center}
\caption{Rest-mass density $\rho$ (normalized to its initial value $\rho_0$) on the $xy$ (top panels) and $xz$ planes (bottom panels) for the B0S3 configuration ($a=0.3, \beta_0^{-1}=0$). The snapshots were taken, respectively, after $\sim$1 orbit, after $\sim$8 orbits and after a time equal to $\sim300 \ M$ after the merger. The regions inside the BH horizons have been masked out.}\label{fig:rhoevolution-B0}
\end{figure*}
In physical units, the adimensional ratios \eqref{eq:beta0} and \eqref{eq:zeta} are
\begin{equation}\label{eq:beta0-2}
    \beta^{-1}_0 \equiv \frac{p_{\mathrm{mag}}}{p_{\mathrm{fluid}}} = \frac{B_0^2}{8\pi \kappa\rho_0^{\Gamma}}
\end{equation}
\begin{equation}\label{eq:zeta-2}
    \zeta_0 \equiv \frac{u_{\mathrm{mag}}}{u_{\mathrm{fluid}}} = \frac{B_0^2}{8\pi c^2\rho_0}
\end{equation}
Therefore, given a specific value of the adimensional parameter $\zeta_0$, one has
\begin{equation}\label{eq:BfromZeta}
    B_0 = \sqrt{8\pi c^2\rho_0\zeta_0} = \left\{\begin{array}{lr}
         0 \ \ \ \ \ \ \ \ \ \ \ \ \ \ \ \ \ \ \ &\text{(B0 runs)}  \\
         3.36\times10^4 \ \mathrm{G} \ \ \ \ &\text{(B1 runs)} \\
         1.20\times10^5 \ \mathrm{G} \ \ \ \ &\text{(B2 runs)}
    \end{array}\right.
\end{equation}
To calculate the initial physical temperature of the accretion flow we use the equation
\begin{equation}\label{eq:A-temp}
 T_0 = \frac{\mu m_{p}}{k_{B}} \frac{p_0}{\rho_0}=\frac{\mu m_{p}}{k_{B}} \kappa \rho_0^{\Gamma-1}
\end{equation}
where $\mu$ is the mean molecular weight, $m_p$ is the proton mass, $\kappa$ is the polytropic constant and $k_B$ is the Boltzmann constant. In code units, we set $\kappa = 0.2$ (conforming with \citeAliasTwo{Giacomazzo-2012}{Gi12} and  \citeAliasTwo{Kelly-2017}{Ke17}).
To assign the physical value of $\kappa$ in cgs units (which, for a $\Gamma = 4/3$ polytrope, has the dimensions of g$^{-1/3}$cm$^3$s$^{-2}$) we proceed as follows: combining Eqs. \eqref{eq:beta0-2} and \eqref{eq:zeta-2}, we find 
\begin{equation}\label{eq:kappa}
    \kappa = \frac{\zeta_0}{\beta_0^{-1}} c^2 \rho_0^{1-\Gamma}
\end{equation}
Inserting the values in code units for $\kappa, c$ and $\rho_0$ in Eq. \eqref{eq:kappa} yields $\zeta_0/\beta_0^{-1}=0.2$, which is adimensional, and thus independent of the units of measure.
Therefore, entering the cgs values of $c$ and $\rho_0$ in \eqref{eq:kappa}, we get
\begin{equation}\label{eq:kappa-2}
    \kappa \sim 8.34\times 10^{23} \rho_{-11}^{-1/3} \ \  \mathrm{g}^{-1/3}\mathrm{cm}^{3}\mathrm{s}^{-2}
\end{equation}
where $\rho_{-11} \equiv \rho_0/10^{-11} \ \mathrm{g \ cm}^{-3}$.
Employing Eq. \eqref{eq:A-temp} with $\mu=1/2$, $\rho_0=10^{-11}$ g cm$^{-3}$ and $\kappa \sim 8.34\times 10^{23} \  \mathrm{g}^{-1/3}\mathrm{cm}^{3}\mathrm{s}^{-2}$, we find that the initial temperature of the accretion flow is
\begin{equation}
    T_0 \sim 1.1\times 10^{12} \ \mathrm{K}.
\end{equation}
Our physical values of the initial magnetic field magnitudes $B_0$ and initial temperature $T_0$ are consistent with those adopted in other general relativistic simulations of hot accretion flows onto MBHBs, e.g. \cite{Bode-2010, Bode-2012}, \citeAliasTwo{Giacomazzo-2012}{Gi12}, \citeAliasTwo{Kelly-2017}{Ke17}.

\section{Results}\label{sec:results}
With our work we probe the physics of MMAFs onto binaries of spinning BHs, evolving a number of simulations which cover a range of black hole spins and gas magnetization. Following  \citeAliasTwo{Kelly-2017}{Ke17}, we aim at exploring the astrophysical processes which may give rise to electromagnetic counterparts to GWs, by studying the near-zone mechanisms that could drive EM emission.
We investigate the role of the BH spins and of magnetic fields on the gas dynamics, exploring how those parameters affect the rest-mass density evolution, as well as the velocity of the fluid in the vicinity of the binary.

As a channel of EM emission we consider the Poynting flux, which may provide a powerful supply of energy that can be converted to strong EM emission farther from the BHs \cite{BZ-1977, Paschalidis-2015, Ruiz-2016}.

To make contact with the results of \citeAliasTwo{Kelly-2017}{Ke17}, we evolve similar configurations (their canonical configuration is an equal-mass binary with $d_0=14.4M$, $\rho_0=1$, and $\beta_0^{-1}=0.025$, in a polytropic gas with $\Gamma=4/3$ and $\kappa=0.2$).
\begin{figure}
\begin{center} 
\includegraphics[width=8.6cm]{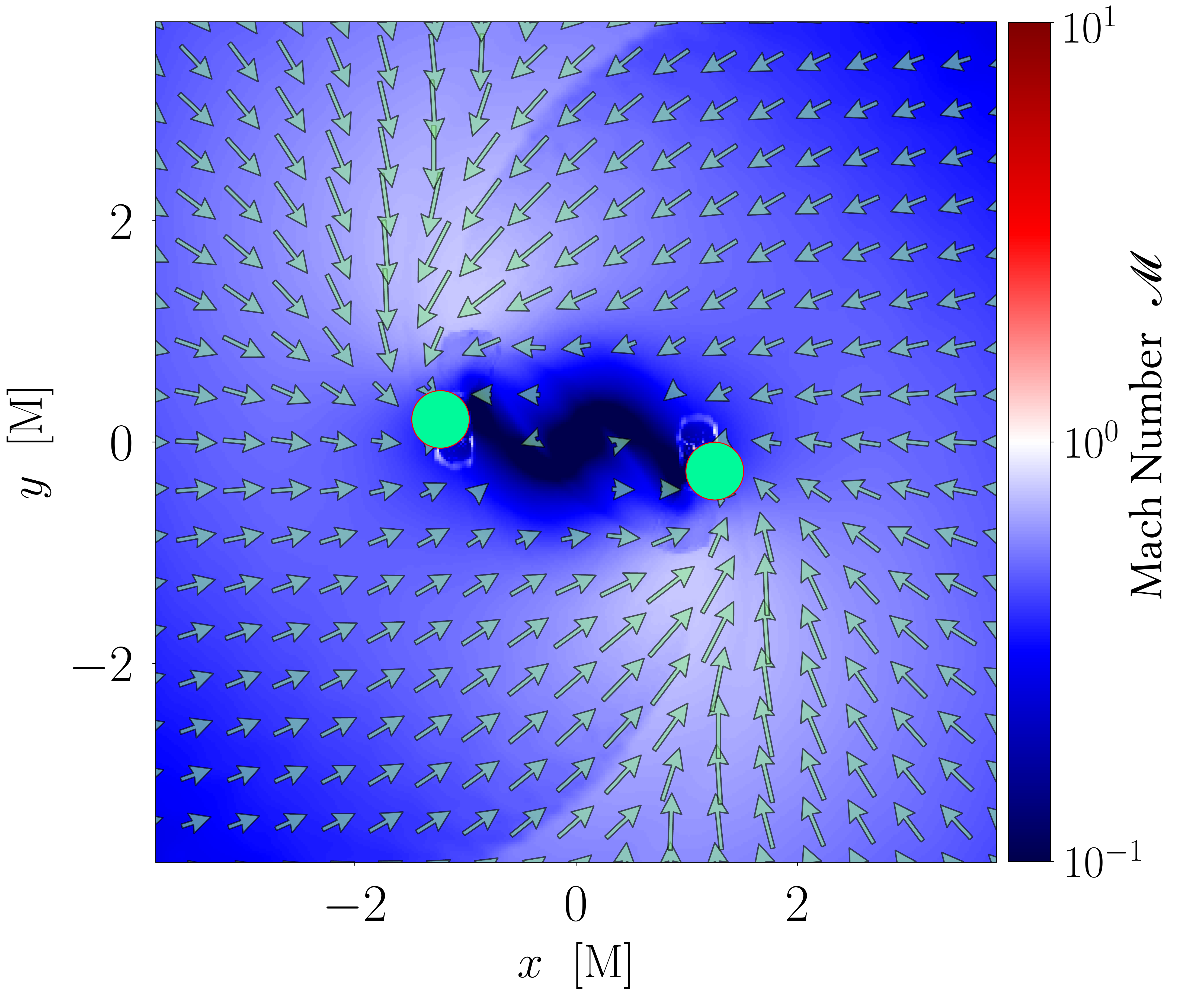}\\
\includegraphics[width=8.6cm]{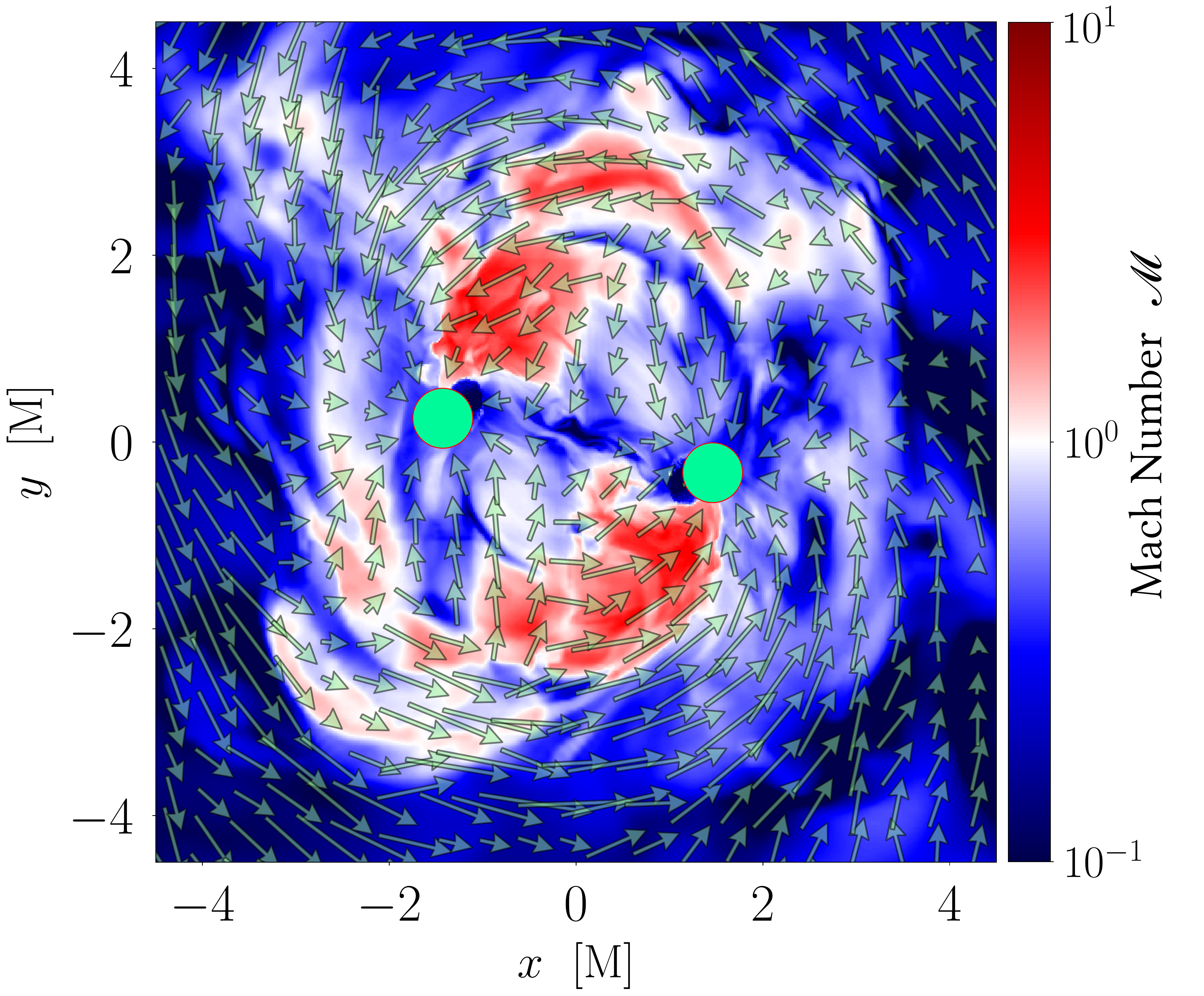}
\caption{Mach number field $\mathscr{M} \equiv \mathrm{v}/c_s$ on the equatorial plane for B0S6 (top) and B2S6 (bottom) models. Arrows denote velocity vectors. The snapshots were taken $\sim1$ orbit before merger. The BHs interiors have been masked out.}\label{fig:BS6Vfield}
\end{center} 
\end{figure}
\begin{figure}
\begin{center} 
\includegraphics[width=8.6cm]{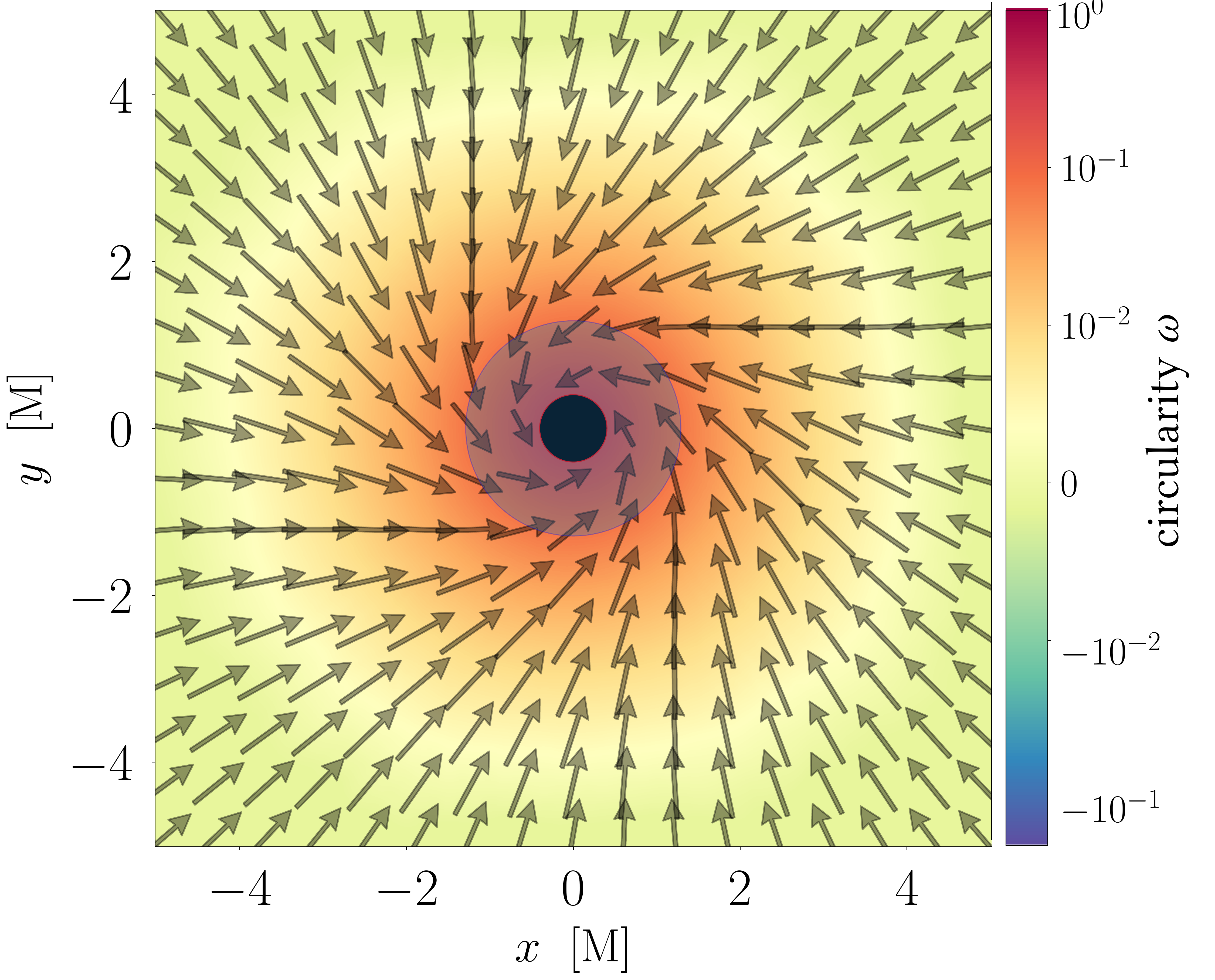}\\
\includegraphics[width=8.6cm]{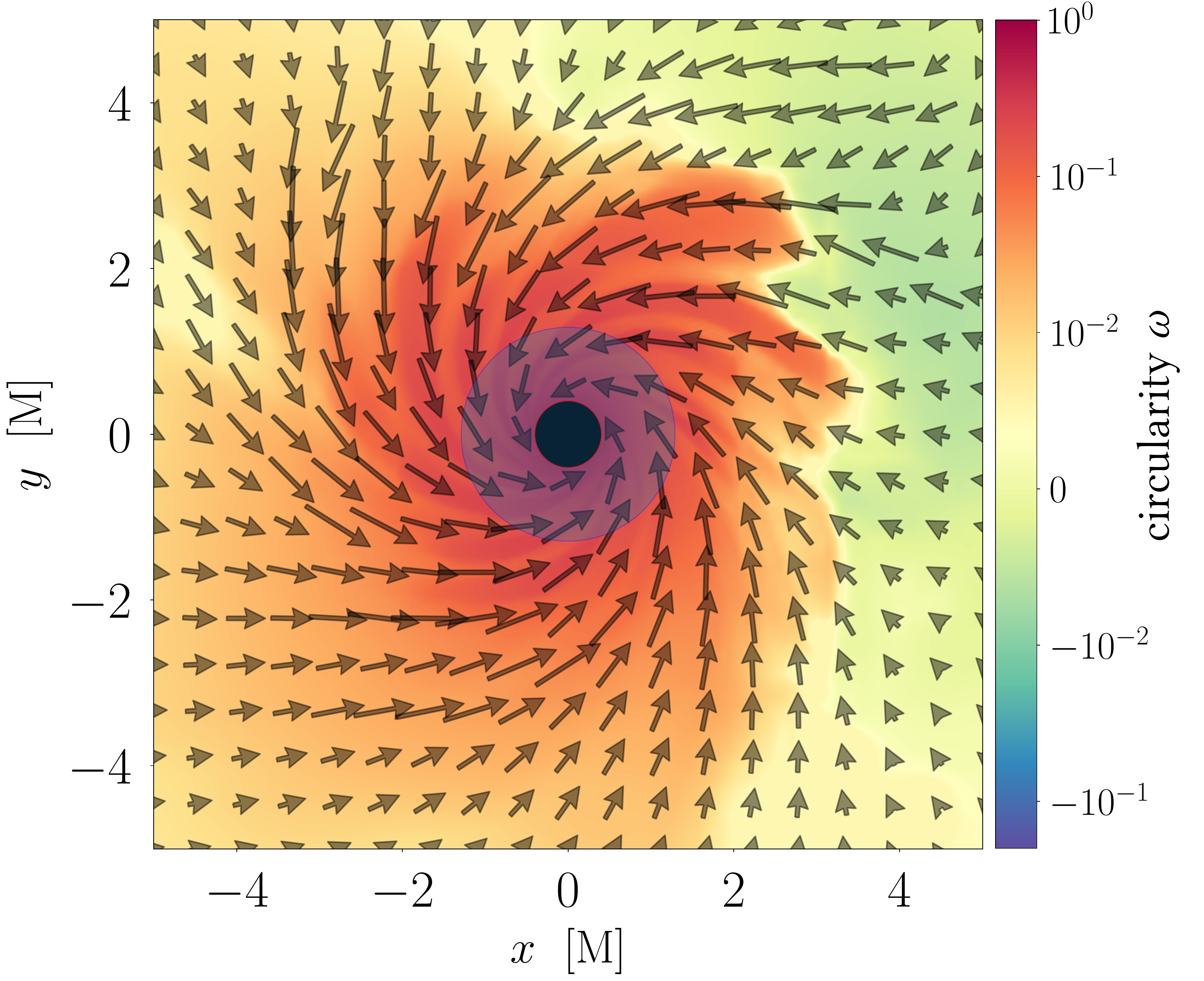}
\caption{Circularity parameter $\omega$ on the equatorial plane for B0S6 (top) and B2S6 (bottom) models. Arrows denote velocity vectors. The snapshots were taken $\sim 100 M$ after merger. The BHs interiors have been masked out. The shaded areas denote the regions within the innermost stable circular orbit ($r < R_{\mathrm{ISCO}}$) for a Kerr BH with spin parameter $a\simeq0.858$ (see, e.g., Eq. (2.21) in \cite{Bardeen-1972}).}\label{fig:BS6OmegaF-Vfield}
\end{center} 
\end{figure}
\subsection{Gas Dynamics}\label{sec:results_gasdyn}
Figures \ref{fig:rhoevolution-B3} and \ref{fig:rhoevolution-B0} show the evolution on the orbital plane $xy$ and on the polar plane $xz$ of the rest-mass density $\rho$ (normalized to its initial value $\rho_0$) for the B2S3 ($\beta_0^{-1} = 0.31$, $a_{1,2}=0.3$) and B0S3 ($\beta_0^{-1} = 0, a_{1,2}=0.3$) configurations. We do not show snapshots for S0, S6 cases since they qualitatively look very similar to S3 models.

The evolution of the unmagnetized model B0S3 (Fig. \ref{fig:rhoevolution-B0}) is similar to the \texttt{B0} configuration (no magnetic fields, nonspinning BHs) in \citeAliasTwo{Giacomazzo-2012}{Gi12}, with the production of two denser gas wakes during the inspiral and the formation of a central spinning BH after merger. Throughout the evolution, the two inspiralling BHs are surrounded by spherical overdensities of matter accreting onto the horizons; after merger, the final BH is ringed by an almost-isotropycal, high-density, spherical distribution of accreting matter (Fig. \ref{fig:rhoevolution-B0}, right column).

The magnetized models exhibit different features (for a comparison, see, e.g., the \texttt{B2} configuration in the work by  \citeAliasTwo{Giacomazzo-2012}{Gi12}, Fig.1, and the \texttt{b1e-1} configuration in \citeAliasTwo{Kelly-2017}{Ke17}, Figs. 3 and 4). In all our magnetized simulations the density close to each BH and in the regions connecting them is larger compared to the unmagnetized cases. We found that the rest-mass overdensities near the BHs in the B1 (B2) models are $\sim$50\% larger than those in the B0 (B1) models; conversely, the individual BH spins show no effect on the enhancement of $\rho$. In the magnetized models, the regions close to the BHs reveal the presence of turbulence in the fluid which is absent in the unmagnetized configurations (see, e.g., the top panels in Fig. \ref{fig:rhoevolution-B3}, which display snapshots of the rest-mass density on the orbital plane $xy$ for the magnetized model B2S3). 
\begin{figure*}
\begin{center} 
\includegraphics[width=5.7333cm]{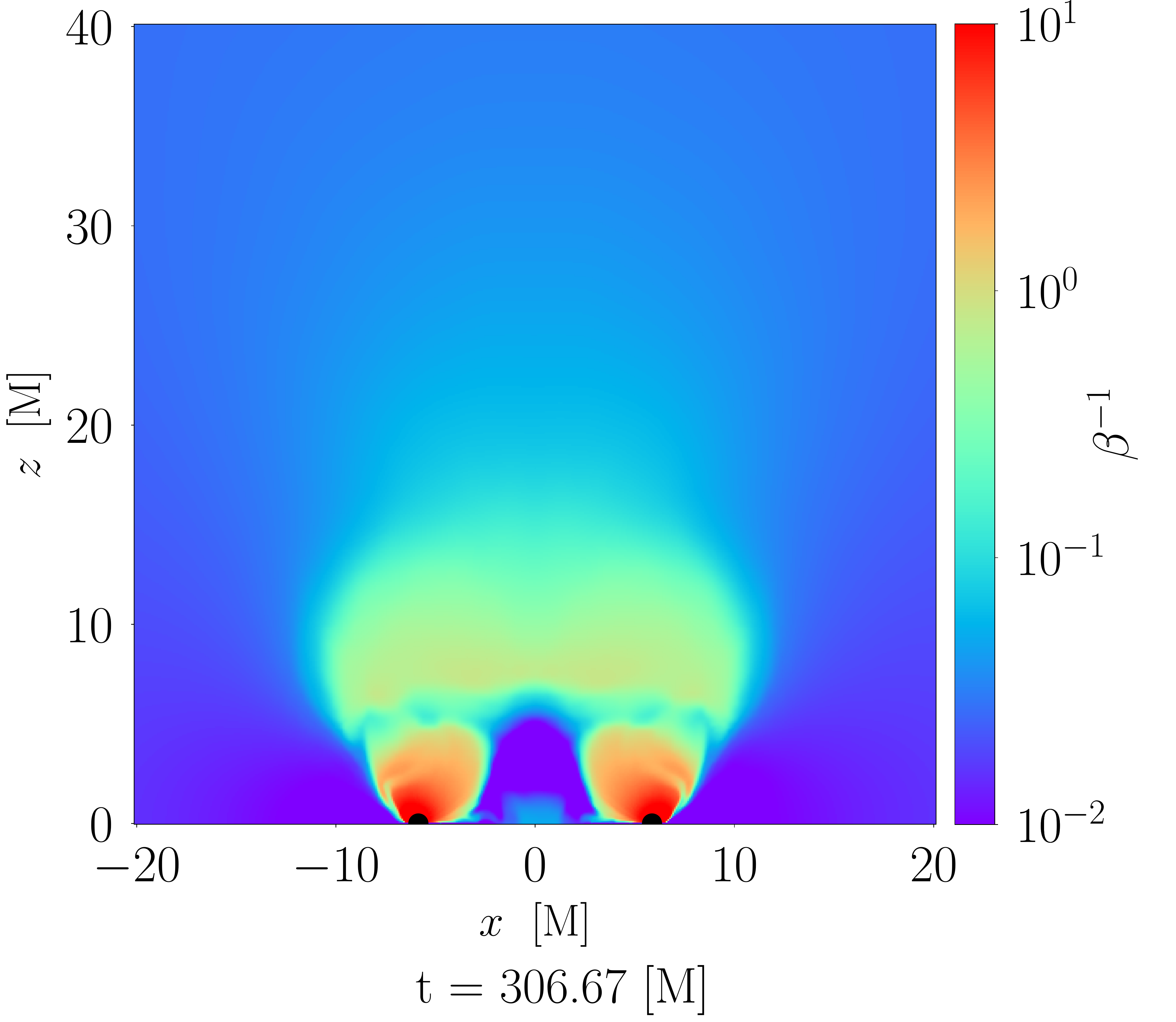}
\includegraphics[width=5.7333cm]{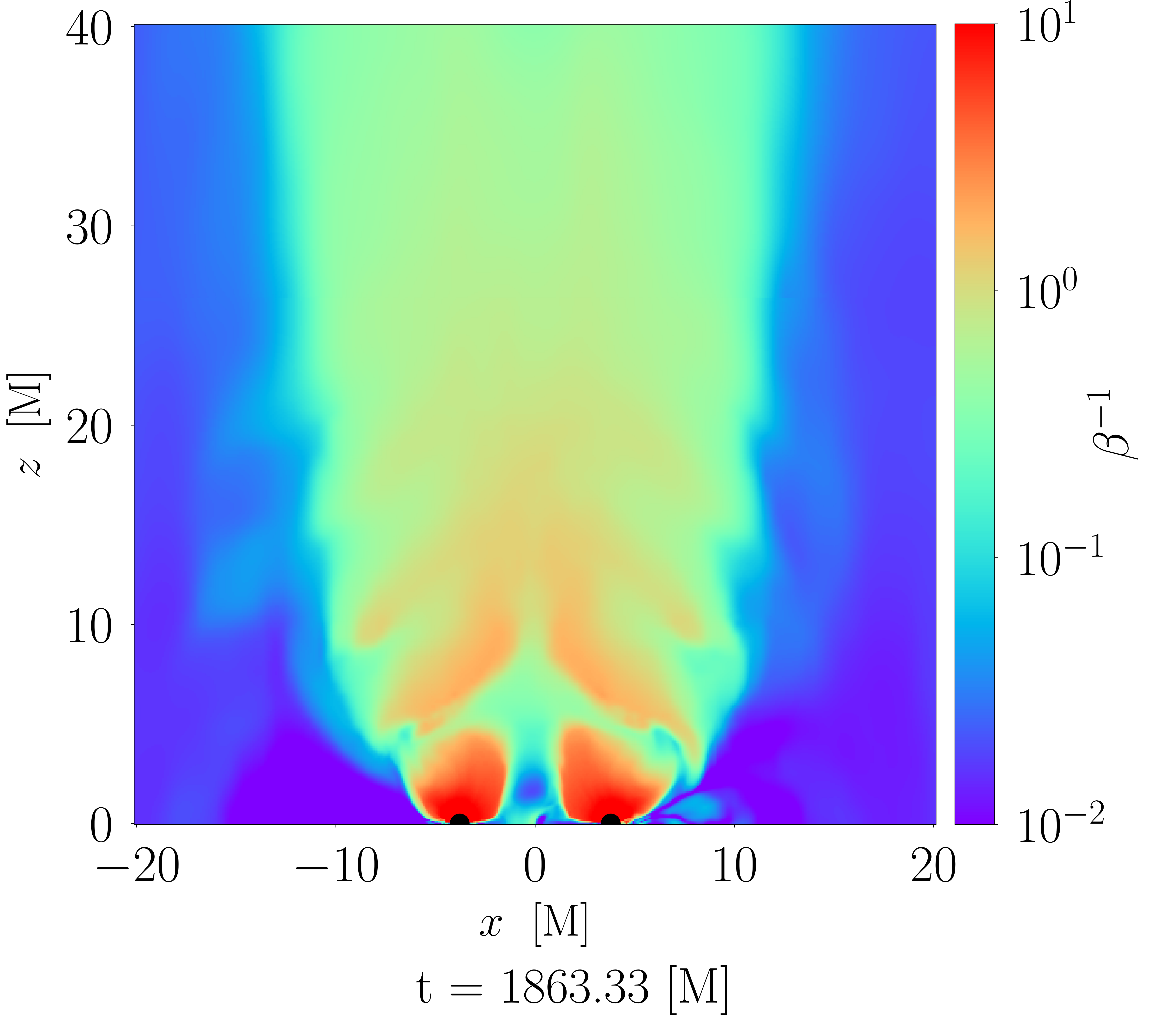}
\includegraphics[width=5.7333cm]{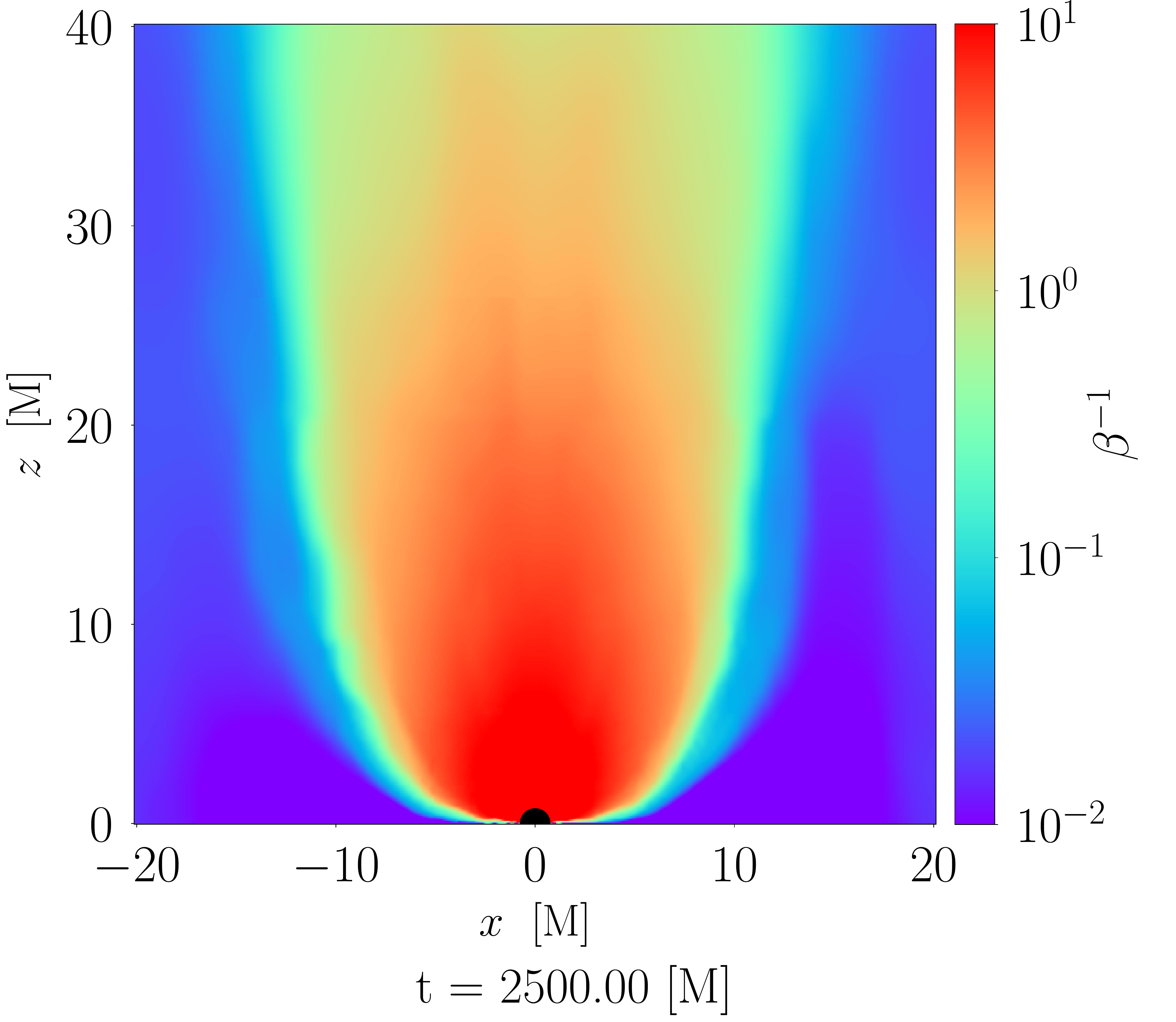}\\
\end{center}
\caption{Evolution of the magnetic to gas pressure $\beta^{-1} = b^2/2p_{\mathrm{fluid}}$ on the $xz$ plane for the B2S3 configuration. The snapshots were taken, respectively, after $\sim$1 orbit, after $\sim$8 orbits and at a time equal to $\sim300 \ M$ after the merger.}\label{fig:beta-evo}
\end{figure*}
\begin{figure}
\begin{center} 
\includegraphics[width=8.6cm]{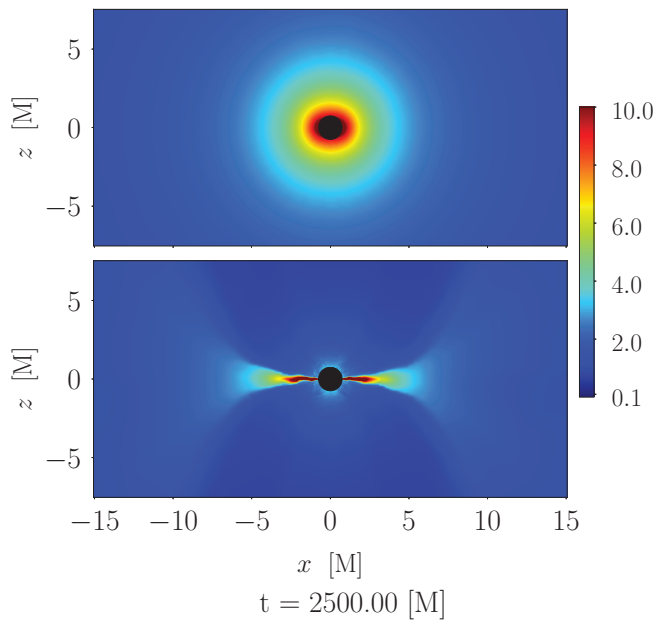}
\caption{Rest-mass density $\rho$ (normalized to its initial value $\rho_0$) on the $xz$ plane for the B0S3 (top panel) and B2S3 (bottom panel) configurations, at $t \sim300 \ M$ after the merger. The regions inside the BH horizons have been masked out.}\label{fig:rho-match}
\end{center} 
\end{figure}
Figures \ref{fig:BS6Vfield}-\ref{fig:BS6OmegaF-Vfield} highlight the differences in the dynamical evolution of the accretion flows between the unmagnetized model B0S6 $(\beta_0^{-1}=0, \ a_{1,2}=0.6)$ and magnetized model B2S6 $(\beta_0^{-1}=0.31, \ a_{1,2}=0.6)$. For each configuration, Fig. \ref{fig:BS6Vfield} displays a two-dimensional snapshot of the Mach number field $\mathscr{M}$ on the equatorial plane $xy$, taken approximately one orbit prior to coalescence. In the unmagnetized case B0S6 (top panel), the fluid is mostly subsonic. The two spiral fronts of the shock waves travel at transonic speed through the inspiral, and are present all the way down to merger. In the magnetized case B2S6 (bottom panel), the shock fronts are hardly visible. The motion of the fluid is more chaotic, and the gas speed in the regions close to the BHs is supersonic. 

Differences between the unmagnetized and magnetized configurations are noticeable also postmerger. In Fig. \ref{fig:BS6OmegaF-Vfield} we show two-dimensional snapshots of the circularity parameter $\omega$ (Eq. \eqref{eq:rotparameter}) for the B0S6 and B2S6 models. Both snapshots were taken $\sim100 M$ after coalescence, and display the magnitudes of $\omega$ and the fluid velocity fields around remnant Kerr BHs with spin parameter $a \simeq 0.858$ \cite[][]{Colpi-Sesana}. The shaded areas mark the regions within the innermost stable circular orbit \cite[][]{Bardeen-1972}. In the unmagnetized model, the accretion flow on the equatorial plane is nearly radial at distances greater than or equal to $3M$, and the circularity parameter at $r_{\mathrm{ISCO}}$ is $<0.1$. Conversely, in the magnetized case the fluid exhibits a higher degree of rotation in the $xy$ plane, and the $\phi$-averaged circularity at $r_{\mathrm{ISCO}}$ is $\sim0.3$.

\subsection{Evolution of Magnetic Fields}\label{sec:results_Bfield}
During the evolution of the B1 and B2 models, the initially weak magnetic fields are dragged along each BH, and soon become dynamically important in the polar regions close to the horizons (see Fig. \ref{fig:beta-evo}, left panel). The magnetic field lines are twisted and compressed, producing a magnification of the magnetic field strength. After the coalescence, the magnetic field strength in the polar regions surrounding of the remnant BH is amplified by a factor $\sim 10^2$; this amplification is observed in all magnetized configurations (in agreement with \citeAliasTwo{Giacomazzo-2012}{Gi12}), and is little-to-not sensitive to the initial $\beta^{-1}_0$ parameter and to the individual BH spins.

In Fig. \ref{fig:beta-evo} we show the evolution of the magnetic-to-gas pressure ratio $\beta^{-1}$ on the $xz$ plane for the B2S3 model. After as short as one orbit (left panel), we see that the polar regions close to the individual horizons are magnetically dominated (i.e., they have larger values of $\beta^{-1}$ than the initial conditions). After eight orbits (central panel) these regions become more pronounced, and outline two vertical areas which are depleted of gas. After merger, a magnetically dominated funnel is created around the spin axis of the remnant BH (right panel). Along this region, the magnetic field strength is increased by a factor $\sim 10^2$, contributing considerably to the total pressure in the gas.

The effect of the magnetically dominated regions on the plasma distribution is noticeable in Fig. \ref{fig:rho-match}, in which we compare the rest-mass distributions on the $xz$ plane after merger (at the same time $t=2500M$) for the B0S3 (unmagnetized) and B2S3 (magnetized) models: the evolution of B0S3 results in a spherical distribution of matter accreting onto the final BH, whereas the end-point of B2S3 evolution is the formation of a thin, ``disklike'' structure around the BH (see also Figs. \ref{fig:rhoevolution-B3} and \ref{fig:rhoevolution-B0}, right panels).
\begin{figure}
\begin{center} 
\includegraphics[width=8.6cm]{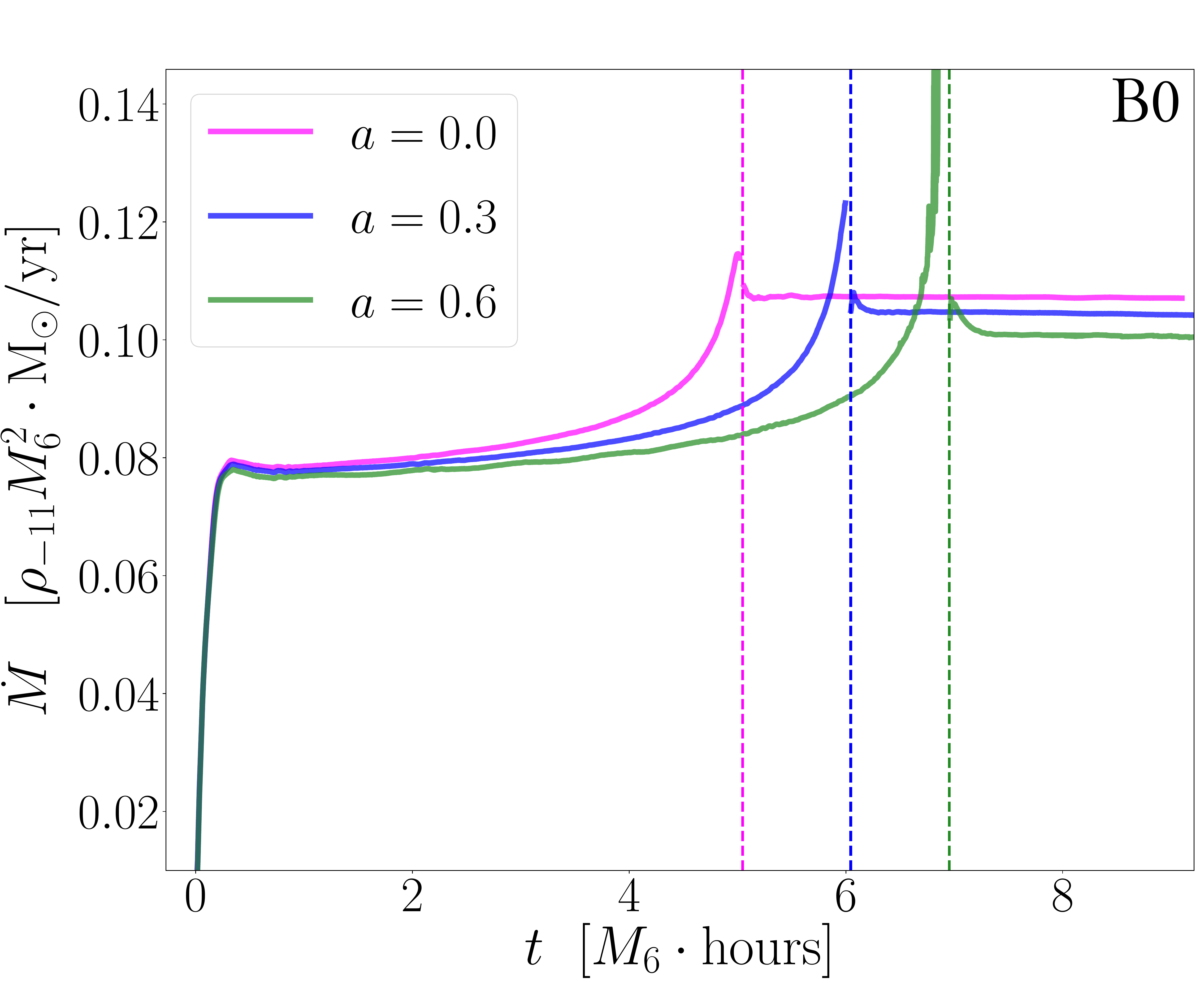}
\end{center}
\caption{Accretion rate $\dot{M}$ in solar masses per year onto the black hole horizons for the unmagnetized (B0) models. The dotted lines mark the merger times for the nonspinning (magenta), $a_{1,2}=0.3$ (blue), $a_{1,2}=0.6$ (green) configurations.}\label{fig:MdotOutflow0}
\end{figure}
\begin{figure}
\begin{center} 
\includegraphics[width=8.6cm]{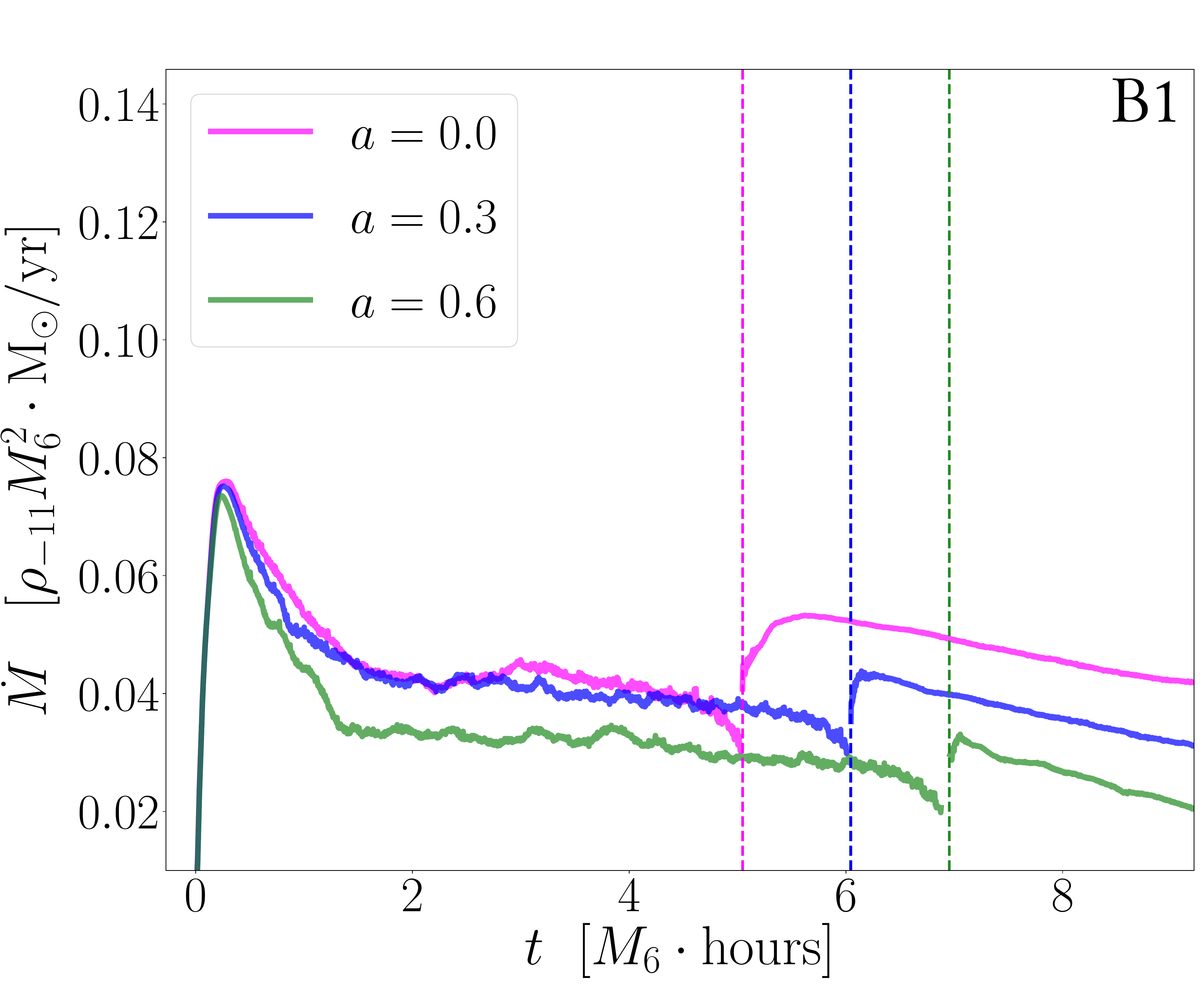}
\end{center}
\caption{Accretion rate $\dot{M}$ in solar masses per year onto the black hole horizons for the  $\beta^{-1}_0=0.025$ (B1) models. The dotted lines mark the merger times for the nonspinning (magenta), $a_{1,2}=0.3$ (blue), $a_{1,2}=0.6$ (green) configurations.}\label{fig:MdotOutflow1}
\end{figure}
\begin{figure}
\begin{center} 
\includegraphics[width=8.6cm]{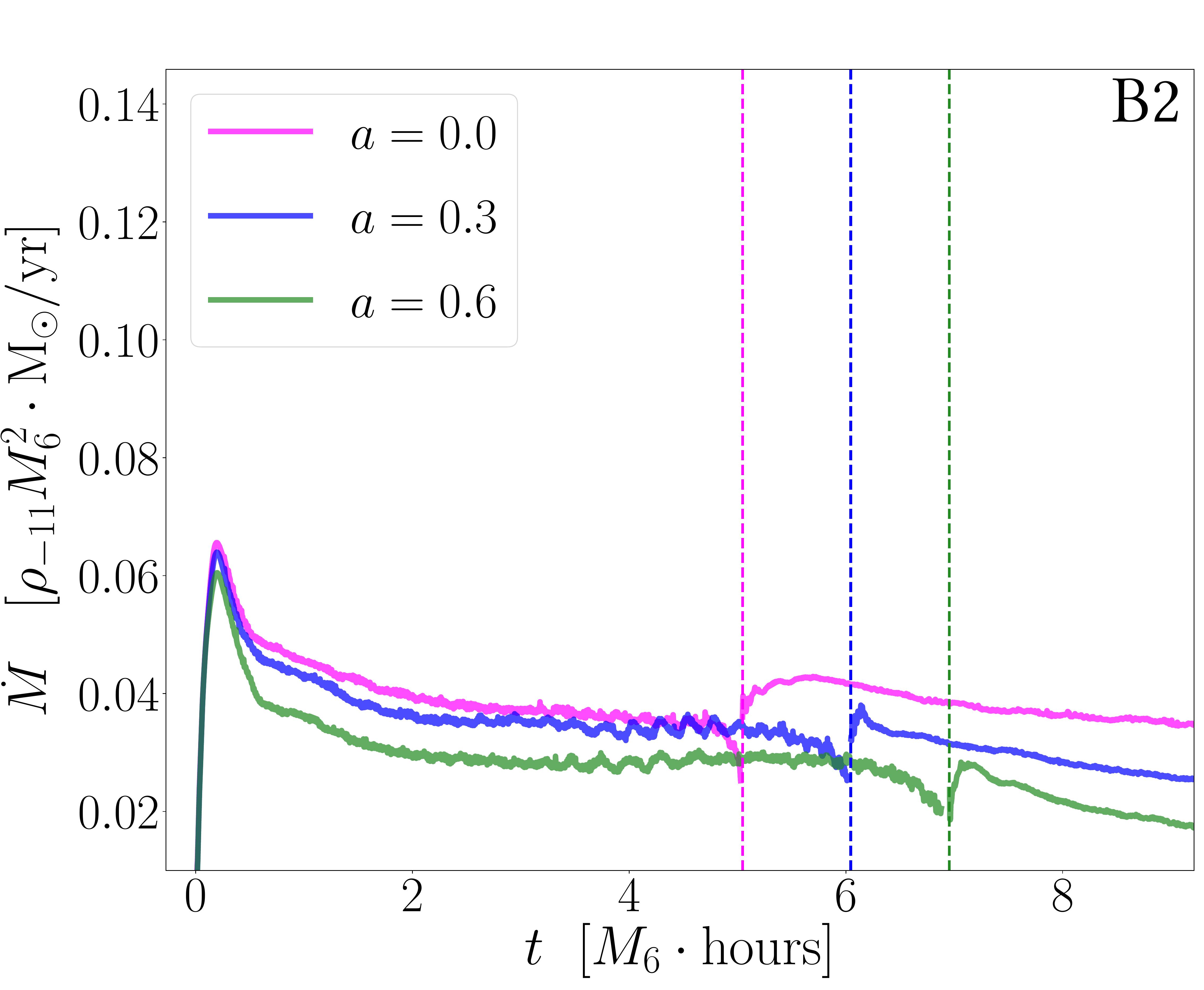}
\end{center}
\caption{Accretion rate $\dot{M}$ in solar masses per year onto the black hole horizons for the  $\beta^{-1}_0 = 0.31$ (B2) models. The dotted lines mark the merger times for the nonspinning (magenta), $a_{1,2}=0.3$ (blue), $a_{1,2}=0.6$ (green) configurations.}\label{fig:MdotOutflow2}
\end{figure}
\begin{figure}
\begin{center} 
\includegraphics[width=8.6cm]{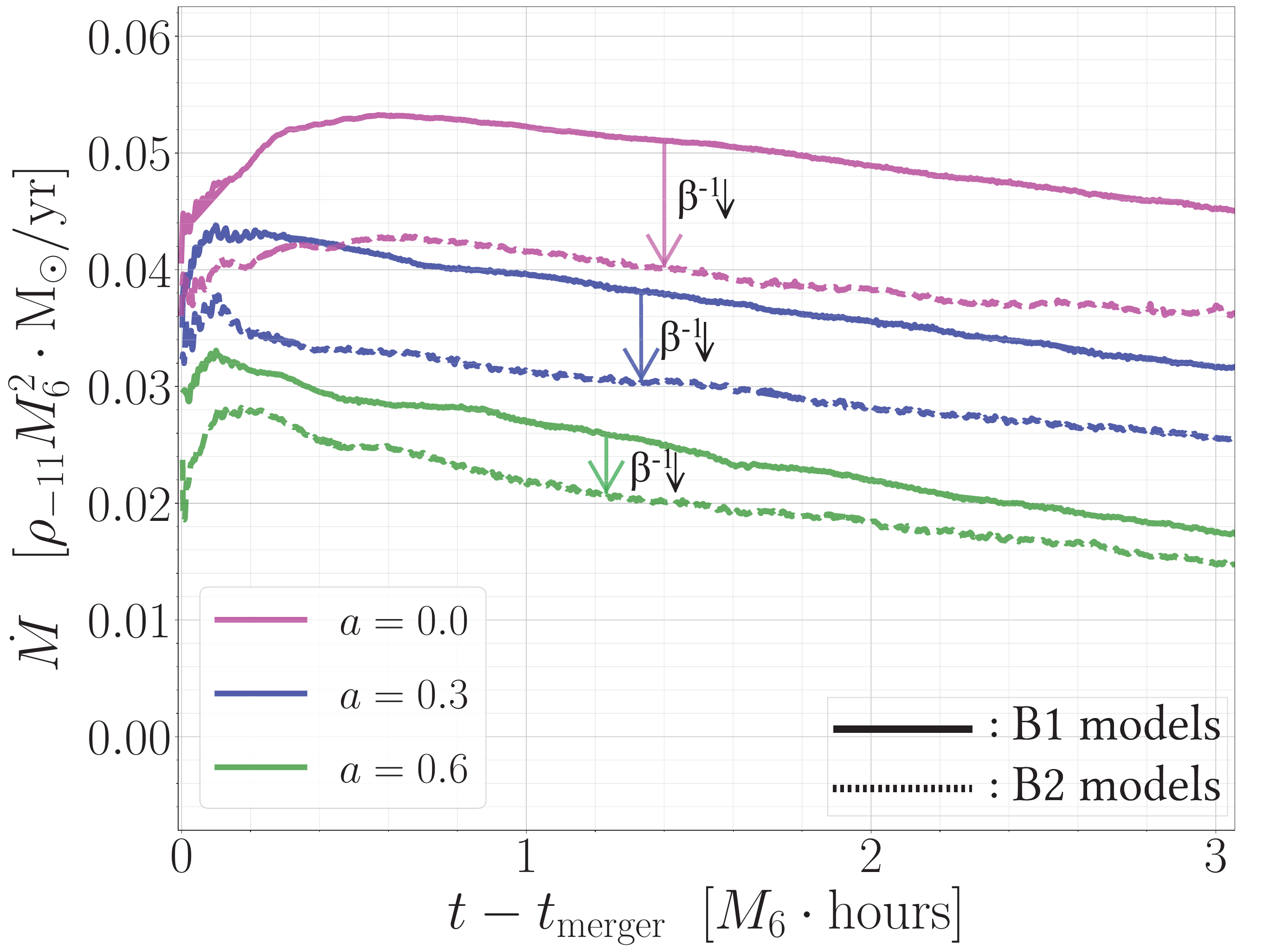}
\end{center}
\caption{Postmerger accretion rate $\dot{M}$ in solar masses per year onto the remnant black hole horizons for the  $\beta^{-1}_0=0.025$ (B1, straight lines) and $\beta^{-1}_0 = 0.31$ (B2, dotted lines) models. The colored arrows denote transitions from lower to higher levels of magnetization $\beta^{-1}$ for same-spin configurations.}\label{fig:Ballout}
\end{figure}

\subsection{Mass Accretion Rate}\label{sec:results_Mdot}
An important diagnostic of our simulations is the flux of rest-mass across the horizons of each BH.
To study the mass accretion rate onto the BH horizons, we use the \texttt{Outflow} thorn \cite[][]{Outflow-thorn}, which computes the flow of rest-mass density across a given spherical surface (e.g., in our case, across each apparent horizon). This quantity is calculated via
\begin{equation}\label{eq:MdotOutflow}
   \dot{M}=-\oint_{S} \sqrt{\gamma}  D v^{i} d \sigma_{i},
\end{equation}
where $D \equiv \rho\alpha u^0$ is the fluid density measured in the observer frame (i.e. $\rho W$, where $W$ is the Lorentz factor), and $\sigma^i$ is the ordinary (flat) space directed area element of the surface enclosing the horizon. Figures \ref{fig:MdotOutflow0}-\ref{fig:MdotOutflow2} show the time evolution of the mass accretion rates $\dot{M}$ onto the BH horizons for each binary system. We plot the evolution of $\dot{M}$ for the unmagnetized (B0) runs, the $\beta^{-1} = 0.025$ (B1) runs, and the $\beta^{-1} = 0.313$ (B2) runs. In each plot (i.e., for each level of magnetization), we compare the values of $\dot{M}$ for the three different spin configurations. The vertical, dotted lines mark the time of coalescence for each spin configuration (as expected, the merger of spinning BHs is delayed as a result of the \textit{hang-up} mechanism \cite[][]{Campanelli-2006b}, which delays or prompts the coalescence according to the sign of the spin-orbit coupling).

The quantities in Fig. \ref{fig:MdotOutflow0}-\ref{fig:MdotOutflow2} are scaled from code to physical units as follows: since $\dot{M}$ generally scales as $\rho M^2$ (g$^3$cm$^{-3}$), we multiply the rate in code units $\dot{M}_{\mathrm{c.u.}}$ by a factor $G^2c^{-3}$ (g$^{-2}$cm$^3$s$^{-1}$) to obtain the rate in cgs units as 
\begin{equation}
        \dot{M}_{\mathrm{cgs}} = \ 6.6\times 10^{21}\dot{M}_{\mathrm{c.u.}} \ \rho_{-11}M_6^2 \ \mathrm{g \ s}^{-1} 
\end{equation}
and the rate in solar masses per year as
\begin{equation}
        \dot{M}_{\mathrm{M}_{\odot}\mathrm{yr}^{-1}} = \ 1.05\times 10^{-4}\dot{M}_{\mathrm{c.u.}} \ \rho_{-11}M_6^2  \ \mathrm{M}_{\odot}\mathrm{yr}^{-1}
\end{equation}
For each level of magnetization, the estimates of $\dot{M}$ share a number of common features:
\begin{itemize}
    \item for the B0 configurations, $\dot{M}$ shows $(i)$ an early maximum as the gas surrounding the binaries establishes a quasiequilibrium flow with the orbital motion, followed by $(ii)$ a steady growth, that reaches its peak at merger $(iii)$. After the coalescence, the accretion rates settle to constant values $(iv)$.
    \item For the B1/B2 configurations, $\dot{M}$ shows the same initial transient as B0  $(i)$, followed by a steep decrease $(ii)$, after which it settles to quasiconstant values which slowly decline prior to merger $(iii)$. Just before the coalescence, the flows drop $(iv)$, and jump upon merger $(v)$ as the apparent horizons join discontinuously.
\end{itemize}
The accretion rates of the magnetized configurations are generally smaller than in the unmagnetized cases by a factor $\sim$2-3. To highlight the effect of different spin parameters on the accretion rate, we show in Fig. \ref{fig:Ballout} the values of $\dot{M}$ for the magnetized models B1 and B2. We focus on the postmerger accretion onto the remnant Kerr BHs. Same-colored lines denote same-spin models, whereas straight (dotted) lines stand for B1 (B2) models. We see that a higher initial magnetization (B2 to B1) has a suppressing effect on $\dot{M}$, which is reduced by $\sim$27\% for spin parameter $a_{1,2} = 0$, by $\sim$20\% for spin parameter $a_{1,2} = 0.3$, and by $\sim$13\% for $a_{1,2}=0.6$.

Conversely, for a given value of $\beta_0^{-1}$, we find that $\dot{M}$ is reduced (compared to the nonspinning case) by $\sim$23\% for $a_{1,2}=0.3$, and by $\sim$48\% for $a_{1,2}=0.6$.
\begin{figure*}
\begin{center} 
\includegraphics[width=5.7333cm]{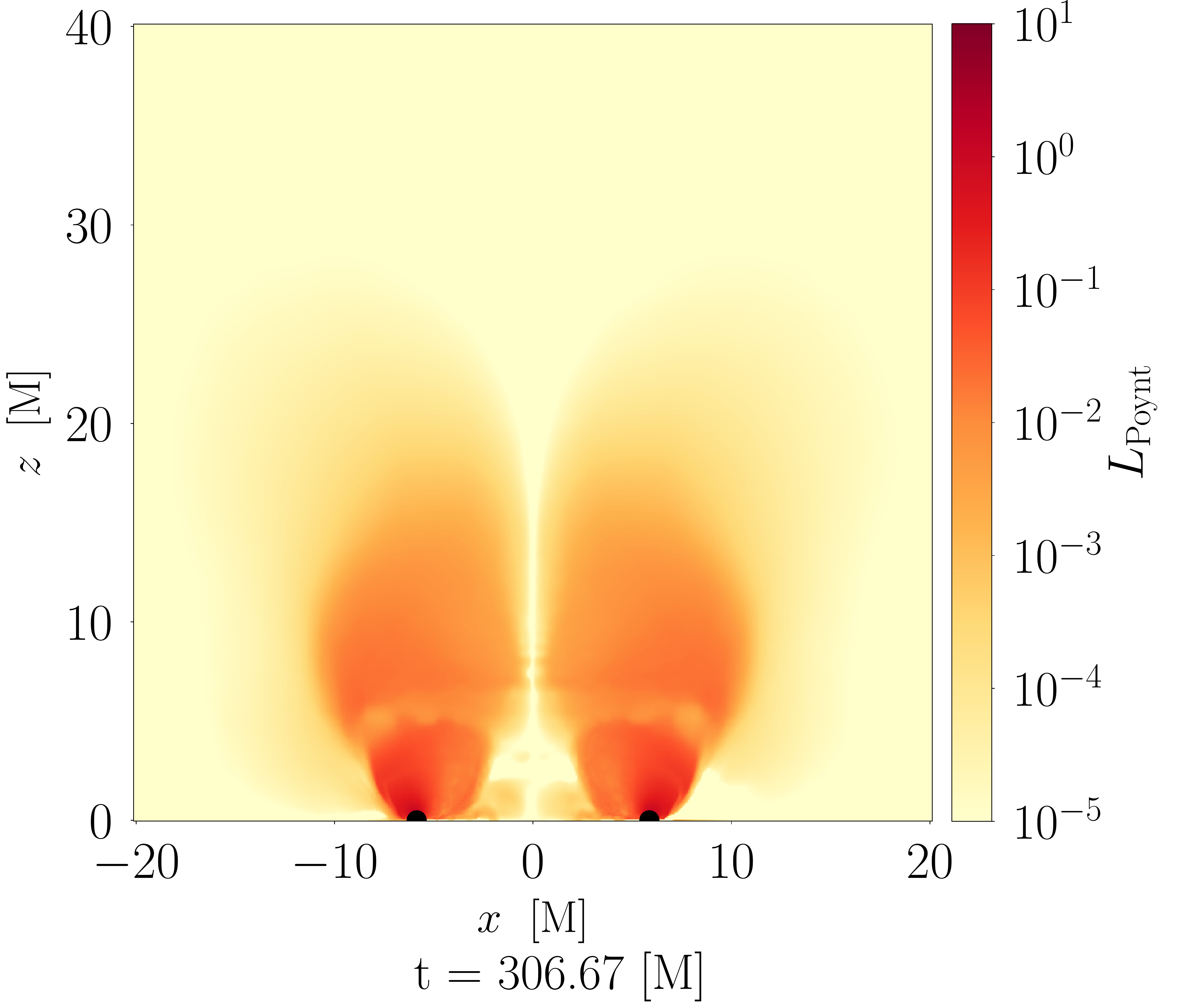}
\includegraphics[width=5.7333cm]{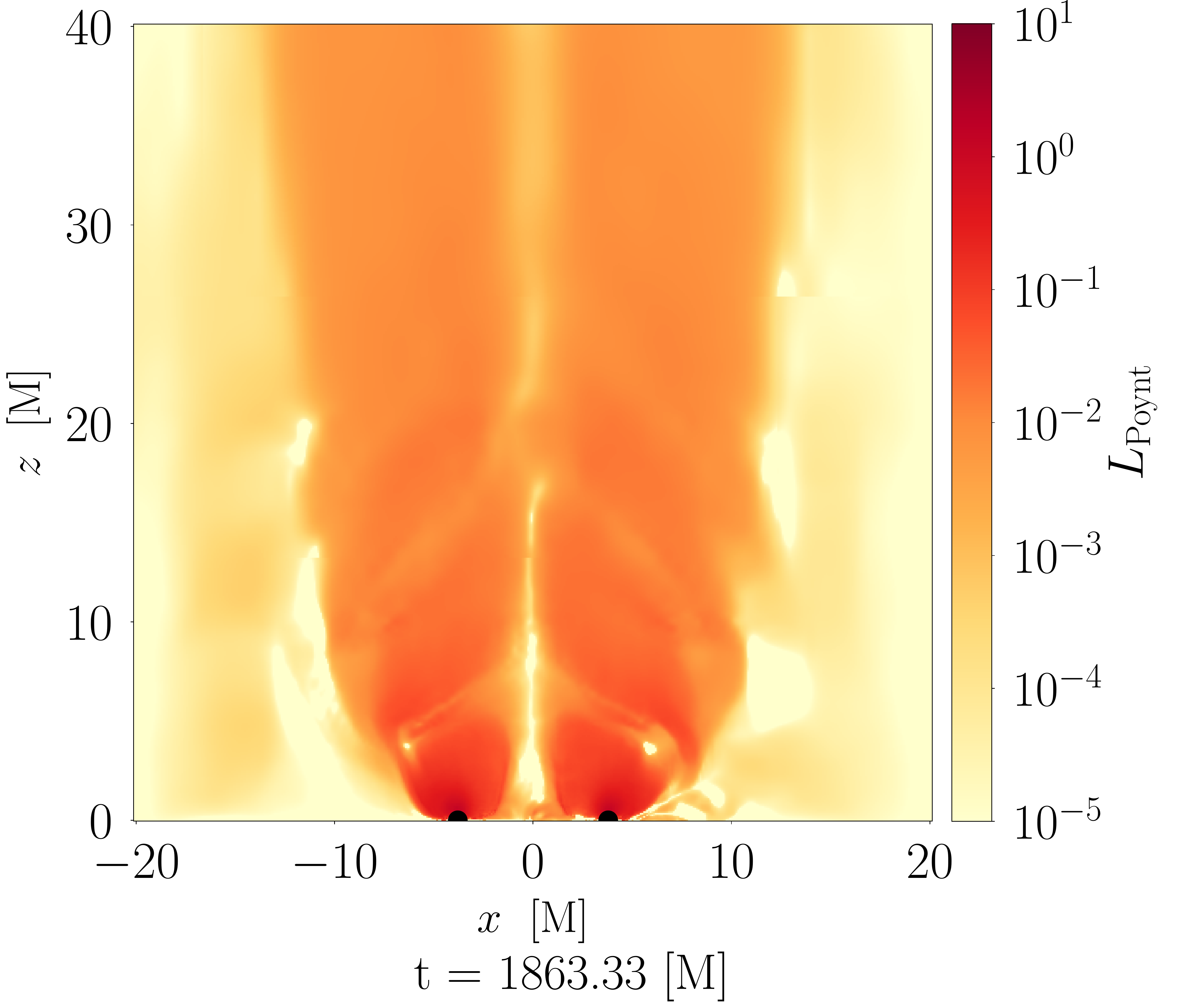}
\includegraphics[width=5.7333cm]{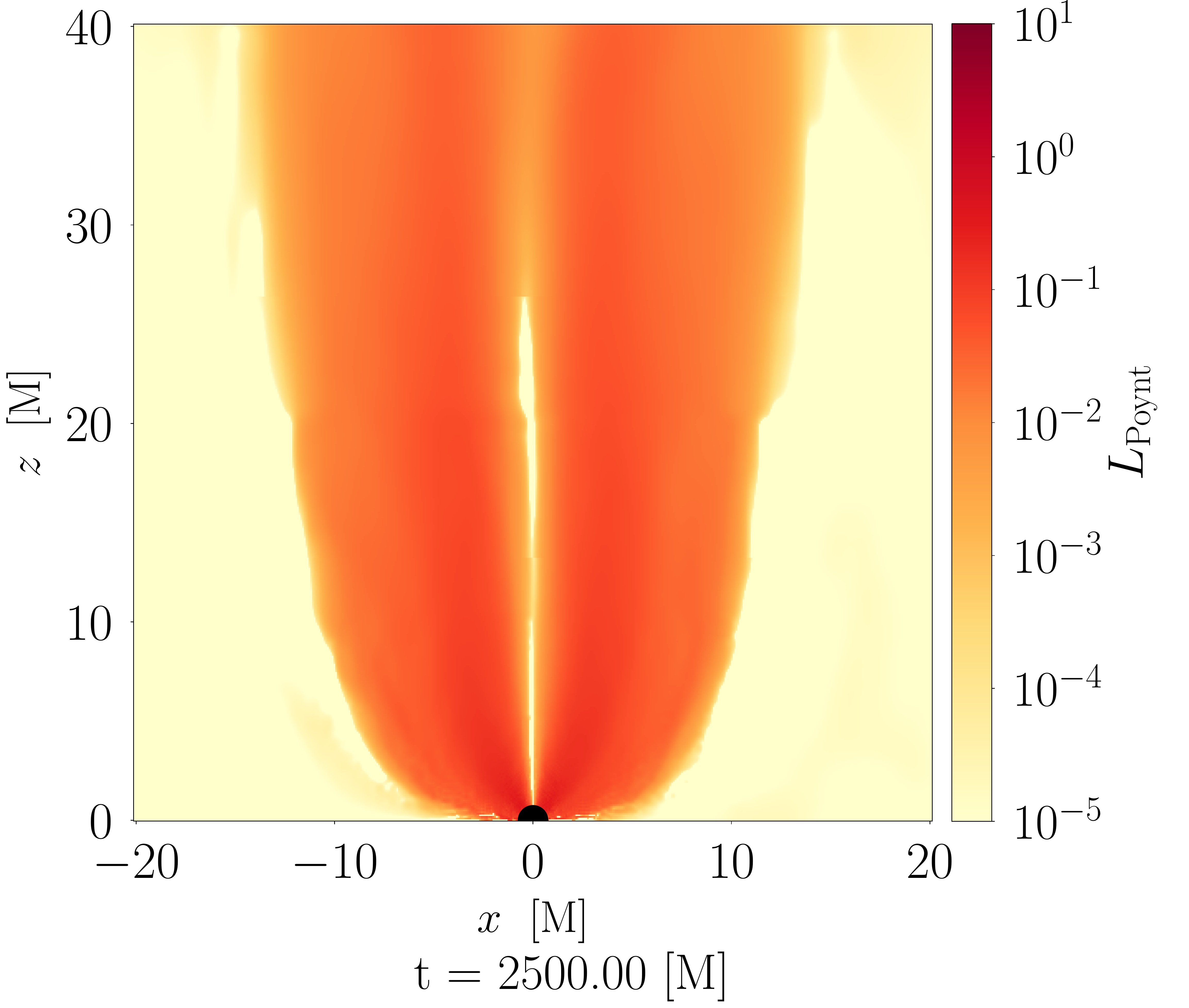}\\
\end{center}
\caption{Evolution of the $z$ component of the Poynting vector (code units) on the $xz$ plane for the B2S3 configuration. The snapshots were taken, respectively, after $\sim$1 orbit, after $\sim$8 orbits and $\sim300 \ M$ after the merger.}\label{fig:poyn-evo}
\end{figure*}
\begin{figure*}
\begin{center} 
\includegraphics[width=8.6cm]{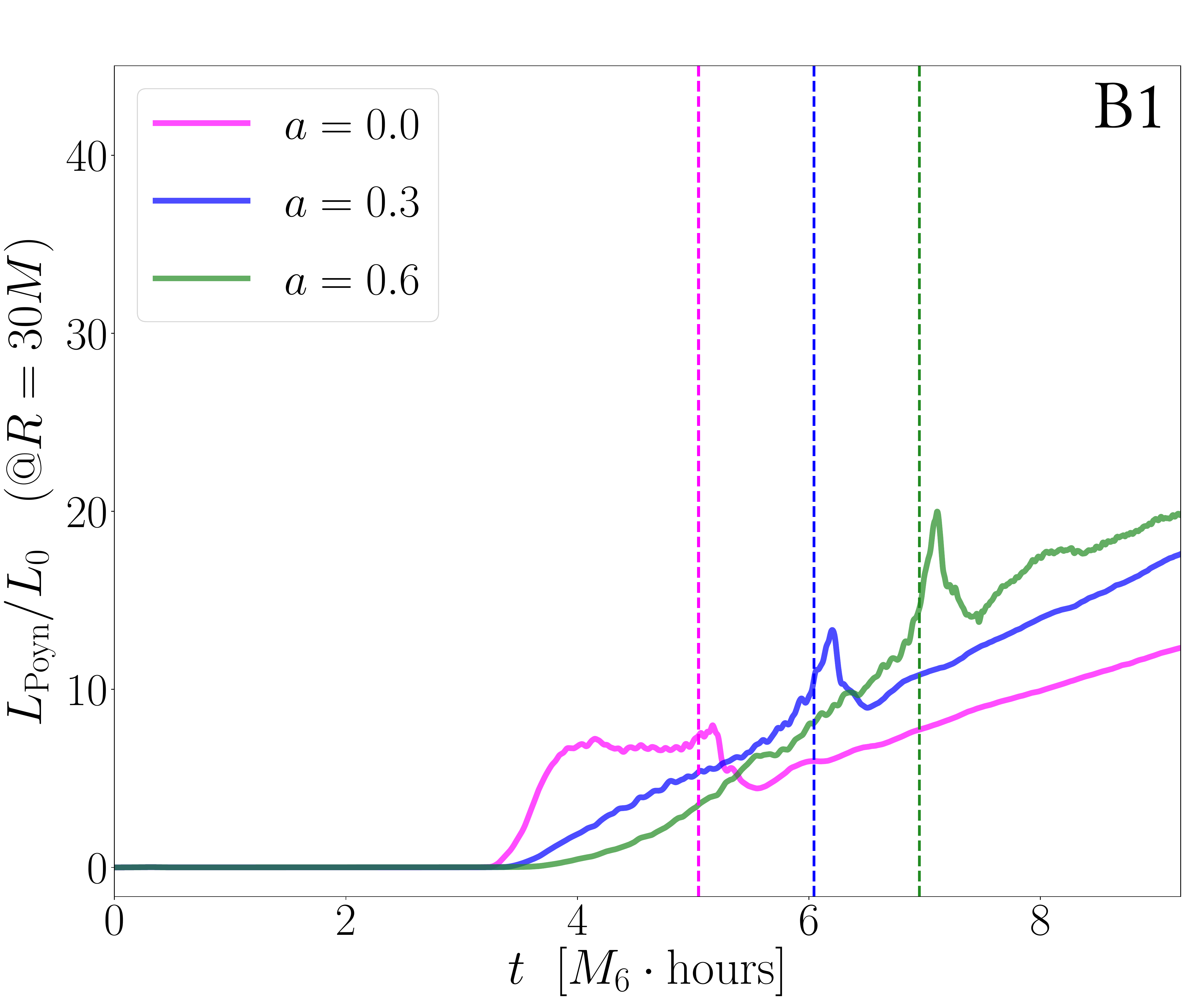}
\includegraphics[width=8.6cm]{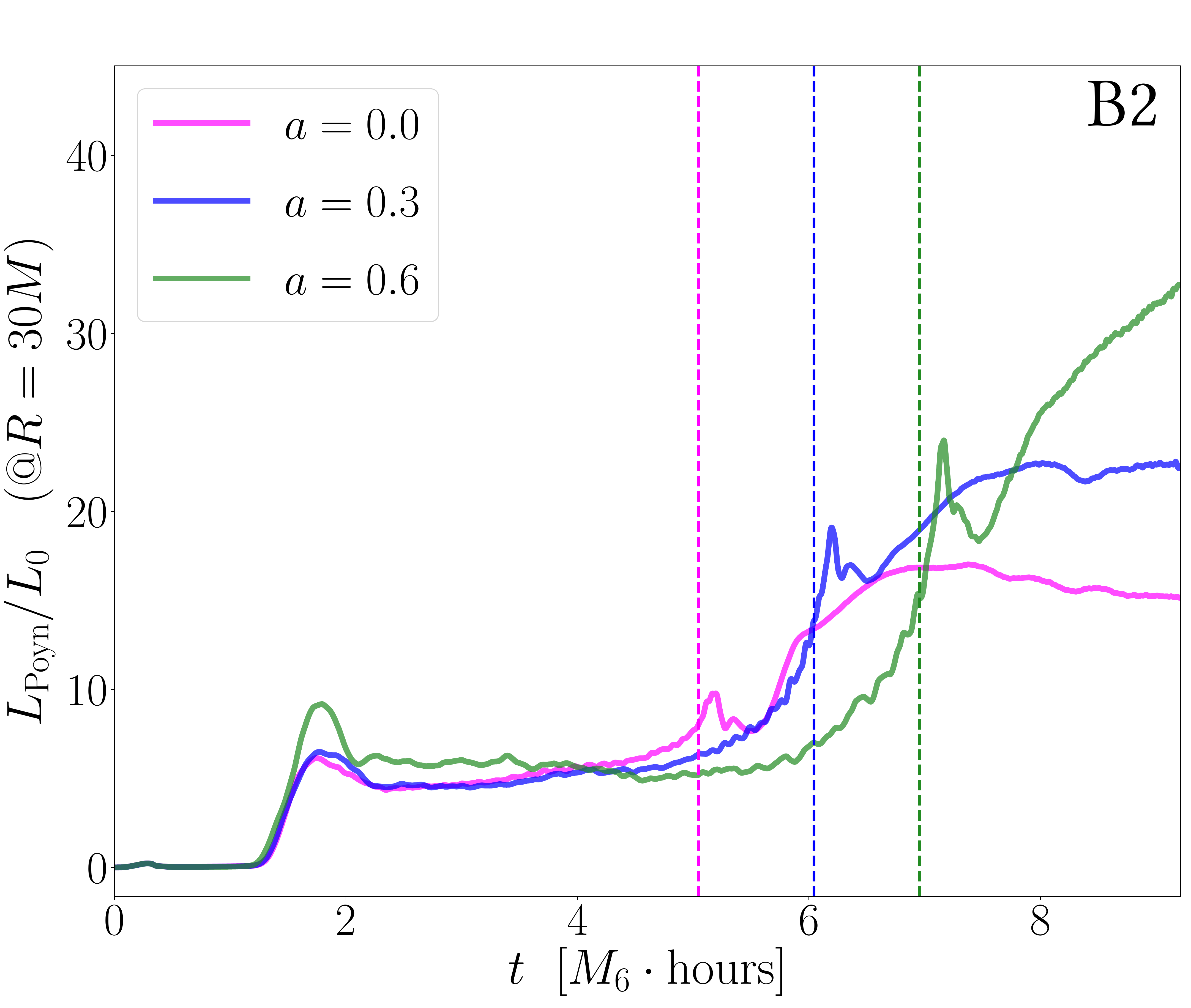}\\
\end{center}
\caption{$L_{\mathrm{Poyn}}$ ($z$ components) in units of $L_0 \equiv 2.347\times 10^{43} \rho_{-11}M_6^2$ erg s$^{-1}$ for the B1 ($\beta^{-1}_0=0.025$, left) and B2 ($\beta^{-1}_0 = 0.31$, right) models, extracted on a coordinate sphere of radius $R=30M$. The dotted lines mark the merger times for the nonspinning (magenta), $a=0.3$ (blue), $a=0.6$ (green) configurations.}\label{fig:Lpoyn}
\end{figure*}

\subsection{Poynting Luminosity}\label{sec:results_poyn}
Several works \cite[][]{Palenzuela-2010, Palenzuela-2010b, Moesta-2010, Moesta-2012} have shown that the interaction of orbiting MBHs with ambient magnetic fields results in the conversion of some of the BH energy into EM energy in the form of collimated regions of Poynting flux. Such regions may generate relativistic outflows \cite[][]{BZ-1977, Paschalidis-2015, Ruiz-2016}, and through a cascade of matter interaction yield strong EM emission.
All our magnetized simulations develop strong flows of electromagnetic energy in the form of Poynting flux; the Poynting flux luminosity can be computed as a surface integral across a two-sphere at a large distance (see Appendix \ref{appendixA}):
\begin{equation}\label{eq:LPoynt-calc}
    L_{\text {Poynt }} \approx \lim _{R \rightarrow \infty} 2 R^{2} \sqrt{\frac{\pi}{3}} S_{(1,0)}^{z}
\end{equation}
where $S_{(1,0)}^{z}$ is the dominant $(l,m)=(1,0)$ spherical mode of the Poynting vector (Eq. \eqref{eq:poynting-def}). Following the evolution of $L_{\mathrm{Poynt}}$ helps us measure the amount of potential emission on timescales comparable to the merger time. To this extent, we extract the luminosity on a coordinate sphere of radius $R_{\mathrm{ext}}$; we set the extraction radius at $R_{\mathrm{ext}}=30M$ as was done in \citeAliasTwo{Kelly-2017}{Ke17} (in \citeAliasTwo{Giacomazzo-2012}{Gi12}, extraction was carried out at $R_{\mathrm{ext}}=10M$, but the initial binary separation was $\sim$30\% smaller than in our simulations). This choice allows us to avoid spurious effects due to the orbital motion of the BHs.

In Fig. \ref{fig:poyn-evo} we show the evolution of the $z$ component of the Poynting vector on the polar plane $xz$ for the B2S3 configuration. As in the simulations of \citeAliasTwo{Giacomazzo-2012}{Gi12}-\citeAliasTwo{Kelly-2017}{Ke17}, the Poynting flux emission in our simulations is largely collimated and parallel to the orbital angular momentum and to the spin of the postmerger BH. In Fig. \ref{fig:Lpoyn} we display the Poynting flux luminosity computed for each of the six magnetized models. On the left, we show the B1 configurations, i.e. those with $\beta_0^{-1} = 0.025$; on the right, we show the B2 configurations, with $\beta_0^{-1} = 0.31$. The values of $L_{\textrm{Poynt}}$ are in units of $L_0 \equiv 2.347\times 10^{43} \rho_{-11}M_6^2$ erg s$^{-1}$ (see Appendix \ref{appendixB}). The values of $L_{\textrm{Poynt}}$ which we observe are consistent with the EM power generated by the Blandford-Znajek \cite{BZ-1977} mechanism \cite[see, e.g., Eq. (4.50) in Ref. ][]{Thorne-1986}:
\begin{equation}
    L_{\mathrm{BZ}} \sim 10^{43} \ \mathrm{erg \ s}^{-1} \left(a\right)^2 \left(\frac{M}{10^6 \ \mathrm{M}_{\odot}}\right)^2 \left(\frac{B}{10^6 \ \mathrm{G}}\right)^2
\end{equation}
The main difference between the B1 and B2 configurations is the time at which the modes reach the extraction sphere at 30$M$. This is what we expected: both configurations evolve a magnetic field which is initially ``dynamically weak'', i.e. the inertia of the plasma is larger than the magnetic field energy. The lower the value of $\beta^{-1}_0$ (B1 configuration), the stronger the initial magnetic field $B_0$ must become in order to surmount the fluid pressure. The development of a stronger magnetic field requires more time; thus, a lower $\beta_0^{-1}$ implies a longer time for launching a jet \cite[][]{Paschalidis-2015}, which is in agreement with our simulations.

We find that the peak luminosity which is reached shortly after merger is sensitive to the BH spins, and is enhanced by a factor of $\sim$2 ($\sim$2.5) for binaries with spin parameters $a=0.3$ ($a=0.6$) with respect to the nonspinning case. This intensification is a general feature, and does not depend on the initial level of magnetization of the gas. Nevertheless, the qualitative behavior of all models is very similar.

\section{Conclusions}
To expand our understanding of the physical processes which arise in the vicinity of merging massive black hole binaries, we carry out GRMHD simulations of equal-mass, spinning MBHB mergers in hot, magnetized environments. We have for the first time investigated the role of individual BH spins in the evolution of magnetic fields and gas dynamics. We evolve a set of nine simulations covering a range of initially uniform, moderately magnetized fluids with different initial magnetic-to-gas pressure ratios. For each magnetization level, we study distinct spin configurations defined by adimensional spin parameters $a=(0, \ 0.3, \ 0.6)$.

Our results offer some insight on the role of spin and magnetization in the magnetohydrodynamical properties of hot accretion flows around merging MBHBs, and on the physical mechanisms which may provide electromagnetic counterparts to future LISA observations.
We have shown that across the orbital evolution, the magnetic field can be distorted by the motion of the BHs and significantly increase its strength, developing magnetically dominated structures in the polar regions above each BH, and ultimately producing a magnetically dominated funnel around the spin axis of the remnant BH.
In general, the dynamics of a magnetized fluid is different than in the unmagnetized case, even if the fluid is initially not magnetically dominated. The accretion flow in magnetized environments yields turbulent motion in the gas near the inspiralling BHs, eventually leading to the formation of a thin, disklike structure rotating on the equatorial plane of the remnant Kerr BH.
These results are consistent with  previous simulations of nonspinning binaries.

We find that mass accretion rates onto BH horizons in magnetized fluids are generally smaller than in unmagnetized cases by a factor $\sim$2-3. For a given initial magnetization, we show that (aligned) spins of the individual BHs have a suppressing effect on the accretion rate as large as $\sim$48\%.

As a potential driver for EM emission, we examined the development and evolution of the Poynting flux. Though not directly observable, it can be considered as a  source of power for EM emission along the jet, and its increase during postmerger evolution may provide observational signs of a merged MBHB. We find that spin can affect the peak luminosity reached shortly after merger, which is enhanced by up to a factor of $\sim$2.5 for binaries of spinning BHs compared to the nonspinning models. This intensification does not depend on the initial level of magnetization $\beta_0^{-1}$ of the gas.

Technical limitations in our analysis prevent more detailed  predictions. Our simulations do not account for the emission of radiation by the plasma, which would determine the magnitude of the accretion luminosity and the shape of the spectra.
Therefore, the accretion flows that we describe lack any radiative mechanism, including cooling and feedback.

In this paper we made an attempt to extract physically relevant information by evolving our simulations in simple gaseous environments, which help us highlight the effects of spins and magnetization on the accretion flows and emitted Poynting luminosity.
While this choice may be useful to identify the subtle effects of different spins and degrees of magnetization, it is not clear how well this simplistic environment can stand in for real accretion flows, which realistically possess angular momentum support and carry dynamical effects from radiation flows. Assessing these limitations motivates our future work.

We aim at extending our exploration of the parameter space of merging MBHBs investigating less symmetrical systems. We will consider binaries with spins which are antialigned with the orbital angular momentum, as well as generic misaligned spin configurations. Also, we intend to study binaries with high-spinning (0.95 and above) MBHs, where one could expect departures from the general trends we found in the present work. Additionally, we will consider binary systems with unequal mass ratios, which are the natural outcome of galaxy mergers in cosmological simulations (as shown, e.g., in Ref. \cite{Volonteri2020}).

These improvements will let us question the effects of surrounding (premerger) material in powering electromagnetic counterparts to gravitational wave events, and the effect of the gravitational recoil imparted to the newly formed MBH on the shock-heated gas along the MBH trajectory.

\begin{acknowledgments}
We thank Bernard Kelly for useful comments and suggestions. 
All simulations were performed on GALILEO and MARCONI machines at CINECA (Bologna,  Italy). Some of the numerical calculations have been made  possible through a CINECA-INFN agreement, providing access to resources on MARCONI (allocation INF20\_teongrav). F.C. acknowledges CINECA award under the ISCRA initiative, for the availability of HPC resources on GALILEO (ISCRA-C Project No. HP10CP7PQ1, allocation IsC83\_HIGHSPIN). M.C. and F.H. acknowledge funding from MIUR under the Grant No. PRIN 2017-MB8AEZ.
\end{acknowledgments}

\appendix
\section{Relation between Poynting Vector and EM flux}\label{appendixA}
In the main text, we calculate the Poynting emission through the (1, 0) spherical harmonic $S^z_{(1, 0)}$ of the $z$-component of the Poynting vector. The quantity $S^z_{(1, 0)}$ is closely related to the EM luminosity computed in the pioneering work by Palenzuela et al. (2010) \cite{Palenzuela-2010}, where the emitted luminosity is determined in terms of the outgoing Newman-Penrose radiative scalar $\Phi_2 = F_{\mu\nu}n^{\mu}n^{*\nu}$. The square of $\Phi_2$ is connected to the electromagnetic energy flux: the EM luminosity is given by the integral
\begin{equation}
    L_{\mathrm{EM}} = \frac{dE_{\mathrm{EM}}}{dt} = \lim_{r\to\infty} \oint \frac{r^2}{2\pi} |\Phi_2|^2 d\Omega.
\end{equation}
The quantity $|\Phi_2|^2$ is proportional to the radial component of the Poynting vector. Assuming that $\Phi_2$ is calculated on a Kerr background using the Kinnersley tetrad, we have
(see, e.g., \cite{Teukolsky-1972} and Eq. (5.13) in \cite{Teukolsky-1973})
\begin{equation}
    \frac{dE_{\mathrm{EM}}}{dt} = \lim_{r\to\infty} \oint r^2 T^r_{ \ 0} d\Omega = \lim_{r\to\infty} \oint \frac{r^2}{2\pi} |\Phi_2|^2 d\Omega
\end{equation}
In the 3+1 formulation of space-time, the quantity $T^r_{ \ 0}$ may be expressed as \cite{Komissarov-2004, Alic-2012} (see also Eq. \eqref{eq:poynting-def} in the main text)
\begin{equation}
    T^r_{ \ 0} = -\frac{1}{\alpha} e^{rjk}E_jH_k = \frac{1}{\alpha}S^r,
\end{equation}
where $e^{ijk}=\sqrt{\gamma}\epsilon^{ijk}$ is the Levi-Civita pseudo-tensor associated to the spatial 3-metric $\gamma$. As $r$ converges to the numerical radial coordinate at large distances, we have $\alpha\to 1$, and the emitted Poynting luminosity can thus be expressed as

\begin{equation}\label{eq:A-Lp1}
    L_{\text {Poynt }} \equiv \lim _{r \rightarrow \infty} \oint r^{2} S^{r} d \Omega = \lim _{r \rightarrow \infty} 2\sqrt{\pi}r^{2} S^{r}_{(0, 0)}
\end{equation}
where $S^{r}_{(0, 0)}$ is the $(l, m)=(0, 0)$ spherical mode of $S^r$. To relate this quantity to the dominant (1, 0) spherical harmonic of $S^z$, we assume that the Poynting flux is dominated by emission along the $z$-axis. Then, we can write
\begin{equation}\label{eq:SrapproxSz}
    S^r \sim S^z\cos\theta
\end{equation}
The $(0, 0)$ and $(1, 0)$ spherical harmonics modes of $S^r$ are related by
\begin{equation}\label{eq:Sr00}
    S^r_{(0, 0)} =\frac{S^r_{(1, 0)}}{\sqrt{3}\cos\theta}
\end{equation}
Therefore, combining Eqs. \eqref{eq:A-Lp1} and \eqref{eq:Sr00}, we find
\begin{equation}
    \lim _{R \rightarrow \infty} 2 R^{2} \sqrt{\frac{\pi}{3}} S_{(1,0)}^{z}
\end{equation}
which is formula \eqref{eq:LPoynt-calc} used in our study.

The assumption \eqref{eq:SrapproxSz} is supported by previous simulations of BHBs in gaseous and magnetized environments. For example, \citeAliasTwo{Kelly-2017}{Ke17} compared the evolution of both $S^z_{(1, 0)}$ and $\sqrt{3}S^r_{(0, 0)}$ and found that, except that for some difference in the initial gauge relaxation, the two signals closely agree.

\vspace{.3cm}
\section{Converting $\bold{\textit{L}_{\mathrm{Poynt}}}$ from code to cgs units}\label{appendixB}
In the GRMHD simulations presented here, the Poynting luminosity scales as
\begin{equation}\label{eq:LPscaling1}
    L_{\mathrm{Poynt}} = \rho_0 M^2 F(t/M; \epsilon_0, \zeta_0)
\end{equation}

where $\epsilon_0$ is the initial specific internal energy, $\zeta_0\equiv u_{\mathrm{mag}}/u_{\mathrm{fluid}}$ the initial magnetic-to-fluid energy density ratio and $F$ is a dimensionless function of time (for more details, see Sec. 3 in \citeAliasTwo{Kelly-2017}{Ke17}).

Equation \eqref{eq:LPscaling1} is in code units, where $c=G=1$. To convert this relation to cgs units, we need to multiply by a factor $G^2/c \approx 1.48 \times 10^{-25}$ g$^{-2}$ cm$^{4}$ s$^{-2}$, and we obtain
\begin{equation}
    \begin{aligned}
L_{\text {Poynt }}(t)=& 1.483 \times 10^{-25}\left(\frac{\rho_{0}}{1 \mathrm{~g} \mathrm{~cm}^{-3}}\right)\left(\frac{M}{1 \mathrm{~g}}\right)^{2} \\
& \times F\left(t ; \epsilon_{0}, \zeta_{0}\right) \operatorname{erg} \mathrm{s}^{-1}
\end{aligned}
\end{equation}
If we want to scale with our canonical density $\rho_0=10^{-11}$ g cm$^{-3}$, and for a system of two BHs of $M_1=M_2=10^6$ M$_{\odot}$ (i.e., $M \simeq 3.977 \times 10^{39}$ g), we find
\begin{equation}
\begin{split}
    L_{\text {Poynt }}(t)=& \ 2.347 \times 10^{43} \rho_{-11} M_{6}^{2} F\left(t ; \epsilon_{0}, \zeta_{0}\right) \text { erg } \mathrm{s}^{-1} \\
    =& \ L_0 \ \rho_{-11} M_{6}^{2} F\left(t ; \epsilon_{0}, \zeta_{0}\right) \text { erg } \mathrm{s}^{-1}
\end{split}
\end{equation}
where $\rho_{-11}\equiv \rho_0/(10^{-11} \mathrm{g \ cm}^{-3})$ and $M_{6}\equiv M/(10^6 \ \mathrm{M}_{\odot})$. The quantity $L_0$ is the normalization factor used in Sec. \ref{sec:results_poyn}, Fig. \ref{fig:Lpoyn}.


\bibliographystyle{apsrev4-2}
\bibliography{cattorini2021}

\end{document}